\documentclass[twocolumn,english,aps,manuscript,prl,reprint,longbibliography]{revtex4-1}
\usepackage[T1]{fontenc}
\usepackage[latin9]{inputenc}
\usepackage{amsmath}
\usepackage{amssymb}
\usepackage{graphicx}
\usepackage{natbib}

\makeatletter

\usepackage[usenames,dvipsnames]{color}
 \@ifundefined{textcolor}{}
 {%
   \definecolor{BLACK}{gray}{0}
   \definecolor{WHITE}{gray}{1}
   \definecolor{RED}{rgb}{1,0,0}
   \definecolor{GREEN}{rgb}{0,1,0}
   \definecolor{BLUE}{rgb}{0,0,1}
   \definecolor{CYAN}{cmyk}{1,0,0,0}
   \definecolor{MAGENTA}{cmyk}{0,1,0,0}
   \definecolor{YELLOW}{cmyk}{0,0,1,0}
 }
\definecolor{light-gray}{gray}{0.55}

\usepackage{babel}

\DeclareMathAlphabet{\mathpzc}{OT1}{pzc}{m}{it}
\makeatother

\usepackage{babel}
\begin{document}

\title{Observation of topological transitions in interacting quantum circuits}

\author{P. Roushan$^{1}$}
\thanks{These authors contributed equally to this work.}
\author{C. Neill$^{1}$}
\thanks{These authors contributed equally to this work.}
\author{Yu Chen$^{1}$}
\thanks{These authors contributed equally to this work.}
\author{M. Kolodrubetz$^{2}$}
\author{C. Quintana$^{1}$}
\author{N. Leung$^{1}$}
\author{M. Fang$^{1}$}

\author{R. Barends$^{1}$}
\author{B. Campbell$^{1}$}
\author{Z. Chen$^{1}$}
\author{B. Chiaro$^{1}$}
\author{A. Dunsworth$^{1}$}
\author{E. Jeffrey$^{1}$}
\author{J. Kelly$^{1}$}
\author{A. Megrant$^{1}$}
\author{J. Mutus$^{1}$}
\author{P. O'Malley$^{1}$}
\author{D. Sank$^{1}$}
\author{A. Vainsencher$^{1}$}
\author{J. Wenner$^{1}$}
\author{T. White$^{1}$}
\author{A. Polkovnikov$^{2}$}
\author{A. N. Cleland$^{1}$}
\author{J. M. Martinis$^{1}$}

\email{martinis@physics.ucsb.edu}

\affiliation{$^{1}$Department of Physics, University of California, Santa Barbara,
CA 93106-9530, USA}

\affiliation{$^{2}$Department of Physics, Boston University, Boston,
MA 02215, USA}

\maketitle

\textbf{The discovery of topological phases in condensed matter systems has changed the modern conception of phases of matter\,\cite{vonKlitzing1980,Tsui1982}. The global nature of topological ordering makes these phases robust and hence promising for applications\,\cite{Nayak2006}. However, the non-locality of this ordering makes direct experimental studies an outstanding challenge, even in the simplest model topological systems, and interactions among the constituent particles adds to this challenge. Here we demonstrate a novel dynamical method\,\cite{GirtsevPNAS} to explore topological phases in both interacting and non-interacting systems, by employing the exquisite control afforded by state-of-the-art superconducting quantum circuits. We utilize this method to experimentally explore the well-known Haldane model of topological phase transitions\,\cite{Haldane1988} by directly measuring the topological invariants of the system. We construct the topological phase diagram of this model and visualize the microscopic evolution of states across the phase transition, tasks whose experimental realizations have remained elusive\,\cite{Alba2011,GoldmanPNAS}. Furthermore, we developed a new qubit architecture\,\cite{Geller2014,Yu2014} that allows simultaneous control over every term in a two-qubit Hamiltonian, with which we extend our studies to an interacting Hamiltonian and discover the emergence of an interaction-induced topological phase. Our implementation, involving the measurement of both global and local textures of quantum systems, is close to the original idea of quantum simulation as envisioned by R. Feynman\,\cite{Feynman1982}, where a controllable quantum system is used to investigate otherwise inaccessible quantum phenomena. This approach demonstrates the potential of superconducting qubits for quantum simulation\,\cite{HouckNP,NoriScience} and establishes a powerful platform for the study of topological phases in quantum systems.}

Since the first observations of topological ordering in quantum Hall systems in the 1980s\,\cite{vonKlitzing1980,Tsui1982}, experimental studies of topological phases have been primarily limited to indirect measurements. The non-local nature of topological ordering renders local probes ineffective, and when global probes, such as transport, are used, interpretations\,\cite{Thouless1982} are required to infer topological properties from the measurements. Topological phases are characterized by topological invariants, such as the first Chern number $\mathcal{C}\mathpzc{h}$, whose discrete jumps indicate transitions between different topologically ordered phases\,\cite{WenQFT,Bernevig}. For a quantum system, $\mathcal{C}\mathpzc{h}$ is defined as the integral over a closed manifold $\mathcal{S}$ in the parameter space of the Hamiltonian as

\begin{equation}\label{chern}
\mathcal{C}\mathpzc{h} \equiv \frac{1}{2\pi} \oint_{\mathcal{S}}  {\bf B} \cdot dS,
\end{equation}
where ${\bf B}$ is the Berry curvature \cite{ch3D,supp,Berry1984}.  As illustrated in Fig.\,1 and discussed in the supplement, ${\bf B}$ can be viewed as an effective magnetic field with points of ground state degeneracy acting as its sources, i.e. magnetic monopoles\,\cite{supp,Wilczek}. Using Gauss's law for the Berry curvature\,(magnetic field), $\mathcal{C}\mathpzc{h}$ simply counts the number of degenerate energy eigenvalues\,(magnetic monopoles) enclosed by the parameter manifold \(\mathcal{S}\). $\mathcal{C}\mathpzc{h}$, which is invariant under perturbations to the shape of \(\mathcal{S}\), is a topological number that reflects a property of the manifold of states as a whole and not a local property of parameter space.

\begin{figure}
\begin{centering}
\includegraphics[width=57mm]{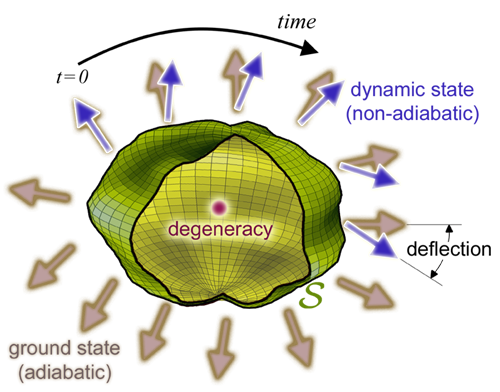}
\caption{
\textbf{Dynamical measurement of Berry curvature and $\mathcal{C}\mathpzc{h}$.} In this schematic drawing, brown arrows represent the ground states (adiabatic limit) for given points on a closed manifold $\mathcal{S}$\,(green enclosure) in the parameter space, and the blue arrows are the measured states during a non-adiabatic passage. According to \eqref{QCF}, the Berry curvature ${\bf B}$ can be calculated from the deviation from adiabaticity. Integrating ${\bf B}$ over $\mathcal{S}$ gives the Chern number $\mathcal{C}\mathpzc{h}$, which corresponds to the total number of degeneracies enclosed.}
\par\end{centering}
\label{fig:concept}
\end{figure}

In previous works, topological properties of highly symmetric quantum systems have been measured\,\cite{NeelyScience,LeekScience,AtalaNP}. However, since these earlier studies relied on interference to evaluate the accumulated phase, these methods are not readily generalizable. To circumvent this, Gritsev et al.\,\cite{GirtsevPNAS} proposed a general method to directly measure the local Berry curvature. The underlying physics of their idea is that motion in a curved space will be deflected from a straight trajectory; in other words, curvature reveals itself as an effective force. For example, a charged particle moving in a magnetic field experiences the well-known Lorentz force. Similarly, Gritsev et al. showed that in a region of the parameter space with Berry curvature ${\bf B}$, if we "move" a quantum system by changing a parameter of its Hamiltonian with rate ${v}$, then the state of  the system feels a force ${\bf F}$ given by
\begin{equation}\label{QCF}
{\bf F} \propto v \times {\bf B}+O(v^2).
\end{equation}
This force leads to deviations of the trajectory from the adiabatic path which can be detected through measurements of the observables in the quantum system\,(see Fig.\,1 and\,\cite{supp}). Therefore, as long as the ramping of parameters is done slowly, but not necessarily adiabatically, the deviation is directly proportional to ${\bf B}$. As the adiabatic limit is generally hard to achieve, this relation has the important advantage of needing only a moderately slow change of state and only requires that the linear term in \eqref{QCF} dominates the dynamics.

This dynamical method suggests a way to directly measure ${\bf B}$, from which $\mathcal{C}\mathpzc{h}$ can be calculated using \eqref{chern}. This provides an alternative means to study topological phases, significantly different from conventional approaches. Admittedly, implementing this procedure requires the ability to continually change the system Hamiltonian, which is difficult to do in most experimental situations. However, in a fully controllable quantum system, this provides a powerful means to probe the topological properties of the ground state manifold through dynamical measurements. Here we demonstrate an implementation of this type of measurement using a quantum circuit based on superconducting qubits\,\cite{Matteo2011,HouckNP,NoriScience}.

\begin{figure}
\begin{centering}
\includegraphics[width=87mm]{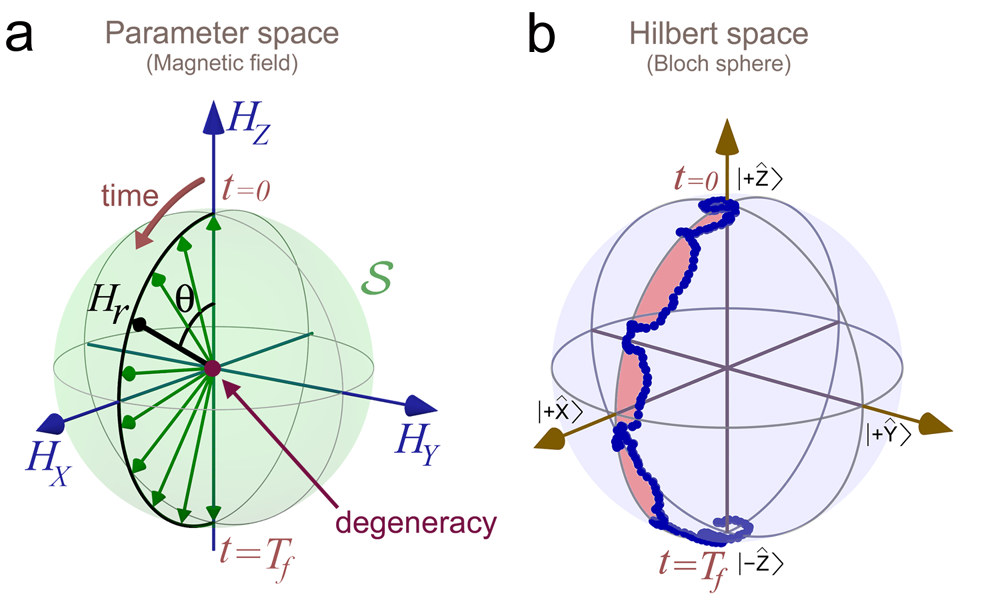}
\par\end{centering}
\caption{
\textbf{Dynamical measurement of $\mathcal{C}\mathpzc{h}$.} \textbf{a.} A simultaneous microwave pulse \(H_X(t)= H_{r}\sin(\pi t/T_f)\) and detuning pulse \(H_Z(t)= H_{r}\cos(\pi t/T_f)\) are applied to construct a parameter space trajectory. The pulse sequence results in a parameter space motion along the \(\phi=0\)  meridian (\(H_{Y}=0\)\,plane) on $\mathcal{S}$. \textbf{b.} The state of the qubit during this ramp (\(H_{r}/2\pi=10\) MHz and \(T_f =600\)\,ns) is determined using tomography\,\cite{supp}, and shown (blue dots) on the surface of the Bloch sphere.}
\label{fig:concept}
\end{figure}

We first demonstrate a basic implementation of the dynamical method. The quantum state of a single superconducting  qubit\,\cite{RamiPRL,supp} is equivalent to a spin-1/2 particle in a magnetic field.  Its Hamiltonian can be written as
\begin{equation}\label{singlemag}
\mathcal{H}_S = -\frac{\hbar}{2}\mathbf{H}\cdot \boldsymbol{\sigma},
\end{equation}
where \(\boldsymbol{\sigma} = (\sigma^x,\sigma^y,\sigma^z)\) are the Pauli matrices, and \(\mathbf{H}=(H_X,H_Y,H_Z)\) is analogous to a control magnetic field.  Full control over the parameters of this Hamiltonian is achieved by microwave pulses that control $H_X$ and $H_Y$, and an applied flux through the qubit's SQUID loop which controls $H_Z$. To illustrate the dynamical method, we measure $\mathcal{C}\mathpzc{h}$ for a spherical ground state manifold in $\mathbf{H}$-parameter space\,(Fig.\,2). We use $\theta$ and $\phi$ as spherical coordinates and consider the parameter trajectory starting at the north pole at $t=0$ and ramps along the $\phi=0$ meridian ($H_Y=0$) with constant velocity $v_\theta=d\theta/dt$  until it reaches the south pole at $t=T_f$.  To realize motion on a spherical manifold, the control sequences of $H_Z$ and $H_X$ are chosen such that the control magnitude $|\mathbf{H}|=H_r$ is constant\,\cite{supp}. In the adiabatic limit, the wavefunction would remain in the instantaneous ground state of $\mathcal{H}_S$, with the Bloch vector parallel to the direction of the control field, following the meridian.  Instead, for non-adiabatic ramps, a deviation from the meridian is observed, as shown in Fig.\,2(b).  Here the Bloch vector is measured at each point in time by interrupting the ramp and performing state tomography.  Note that this deviation is not due to noise, but rather is the expected non-adiabatic response\,\cite{supp}.  For this trajectory, the force ${\bf F}$ takes the form \(f_\phi = \frac{\hbar}{2}H_{r}\left< \sigma^y \right>\sin \theta\), where \(\left<\sigma^y\right>\) is the expectation value of \(\sigma^y\). Integrating over the resulting deflection (shaded light red in Fig. 2(b)) gives $\mathcal{C}\mathpzc{h}=1 \pm 0.05$. Note that given the symmetry of the Hamiltonian, a line integral is sufficient for measuring the surface integral of $\mathcal{C}\mathpzc{h}$ (see \eqref{chern})\,\cite{Schroer2014,supp}. A value of unity is expected, as the qubit ground state has a single degeneracy at $\mathbf{H}=0$, corresponding to an effective monopole, the enclosing parameter sphere $\mathcal{S}$ should yield $\mathcal{C}\mathpzc{h}=1$. In the supplement we demonstrate the robustness of $\mathcal{C}\mathpzc{h}$ by deforming the surface manifold $\mathcal{S}$ and discuss the sources of error,\cite{Vavilov2014}.

\begin{figure}
\begin{centering}
\includegraphics[width=65mm]{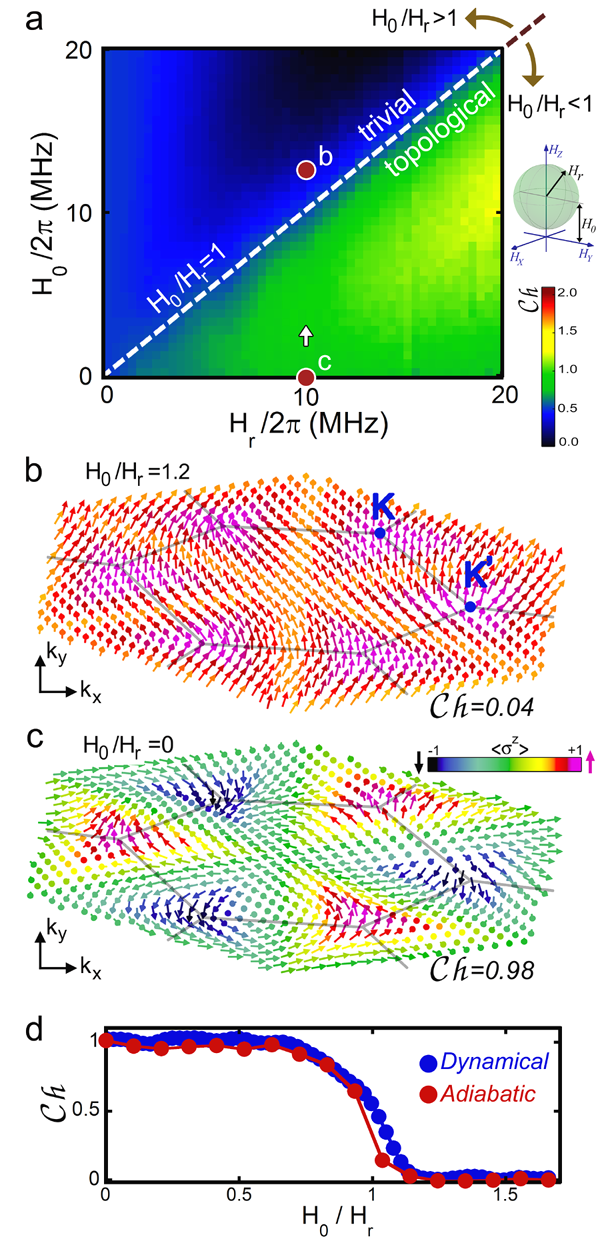}
\par\end{centering}
\caption{
\textbf{Dynamic measurement of the topological phase diagram and adiabatic visualization of phases.} \textbf{a.} Dynamical determination of the phase diagram. first \(\left<\sigma^y\right>\) was measured during ramps similar to those in Fig 2(a), and then $\mathcal{C}\mathpzc{h}$ was calculated. The dashed line is the expected phase boundary at \(H_0=H_r\). \textbf{b},\textbf{c.} With adiabatic state preparation, the state of the qubit was prepared and measured over a grid on the surface of the parameter sphere and then mapped to the hexagonal momentum-space plane. The ground states are presented as Bloch vectors, whose colors indicate their \(\left<\sigma^z\right>\) values. \(H_0/H_r=1.2\) for \textbf{b} and \(H_0/H_r=0\) for \textbf{c}. The gray lines show the FBZ of the honeycomb lattice and high symmetry points \ensuremath{\mathbf{K}} and \ensuremath{\mathbf{K'}} are marked. \textbf{d.} The measured $\mathcal{C}\mathpzc{h}$ from the adiabatic and dynamical (white arrow in \textbf{a}) methods are plotted vs. \(H_0/H_r\).}
\label{fig:concept}
\end{figure}

\begin{figure*}
\begin{centering}
\includegraphics[width=110mm]{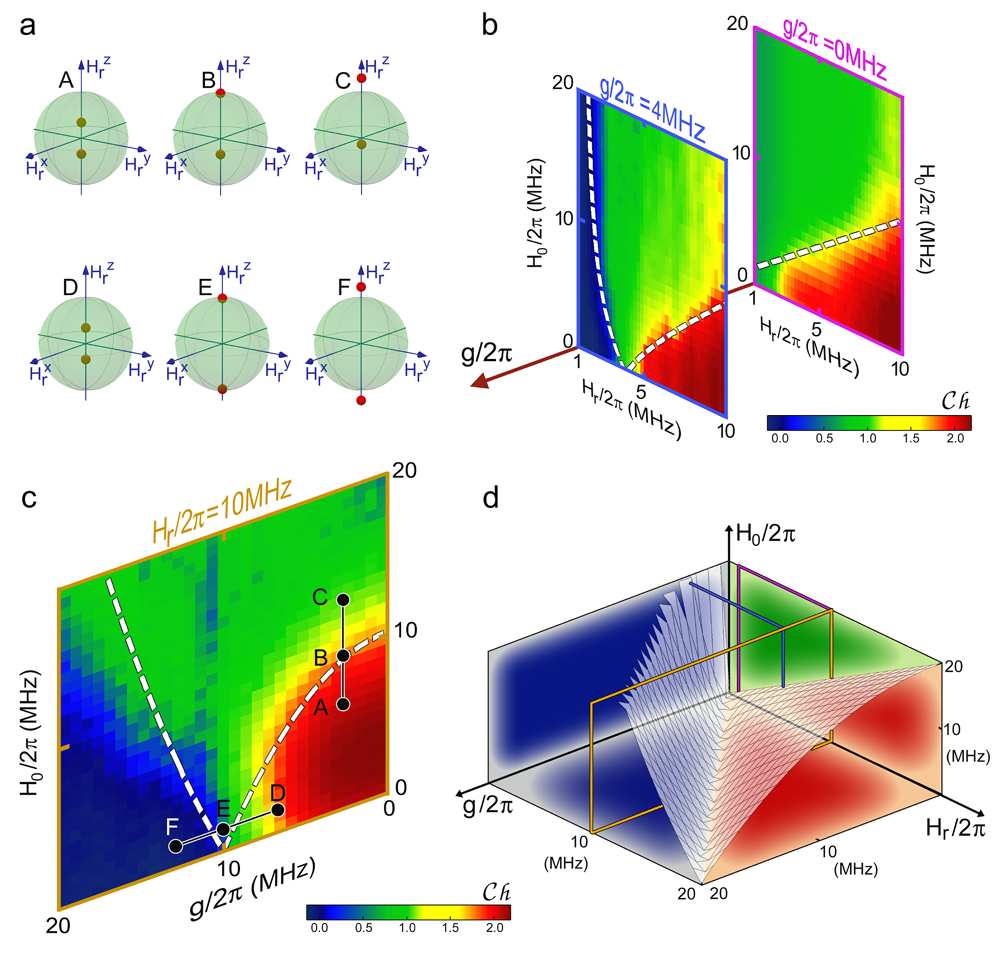}
\par\end{centering}
\caption{
\textbf{Topological phase diagram of an interacting system.}
\textbf{a.} The position of the monopoles in \textit{H}-space for the points A through F shown in panel\,\textbf{c}, with a spherical manifold of radius \(H_{r}/2\pi=10\)\,MHz. \textbf{b,c.} The topological phase diagram of Eq. \eqref{2Q}. In panel \textbf{b}, $\mathcal{C}\mathpzc{h}$ was measured for two fixed \(g/2\pi\) values of 0 and 4 MHz. In panel \textbf{c}, $\mathcal{C}\mathpzc{h}$ was measured for fixed \(H_{r}/2\pi=\)10\,MHz. Dashed lines are topological transitions calculated analytically. \textbf{d.} The analytically calculated phase diagram showing three distinct $\mathcal{C}\mathpzc{h}$ volumes and the separatrix plane. The phase diagram cuts in \textbf{b, c} are indicated by colored slices.}
\label{fig:concept}
\end{figure*}

Using our controllable quantum circuit, we can explore what is perhaps the simplest model of topological behavior in condensed matter, the Haldane model\,\cite{Haldane1988,supp}. This model serves as a foundation for other topological insulator models\,\cite{BHZ,HasanTI,MooreN}, yet its experimental realization has remained elusive\,\cite{Alba2011,GoldmanPNAS}. To show that the quantum Hall effect could be achieved without a global magnetic field, Haldane introduced a non-interacting Hamiltonian on a honeycomb lattice\,\cite{Haldane1988} given by
\begin{equation}
\mathcal{H}_{G}(k_{x},k_{y})=\hbar v_{F}(k_{x}\sigma^{x}+ k_{y}\sigma^{y})+(m_{0}-m_{t})\sigma^{z},\label{graphene}
\end{equation}
where \ensuremath{v_{F}} is the Fermi velocity,\,\ensuremath{m_{0}} is the effective mass, and \ensuremath{m_{t}} corresponds to a second-neighbor hopping in a local magnetic field. The key prediction of the Haldane model is that if \(m_0/m_t>1\)  the system is in a trivial insulating phase, and otherwise in a topological phase, where edge states and quantized conductance appear. Using a confocal mapping\,\cite{supp} one can recast Eq. \eqref{graphene} into the single-qubit Hamiltonian  \eqref{singlemag}. If we consider spherical manifolds $\mathcal{S}$ of radius $H_r$ displaced from the origin in the $z$ direction by $H_0$\,, then \ensuremath{H_{0}/H_{r}} in the qubit system plays the same role as \ensuremath{m_{0}/m_{t}} in the Haldane model. In Fig.\,3(a) we plot the results of this measurement, showing $\mathcal{C}\mathpzc{h}$ as a function of $H_r$ and $H_0$, which shows plateaus at values 0 and 1 separated by a phase transition boundary line at $H_r = H_0$.  This transition can be easily understood: when $H_0 < H_r$ the degeneracy at $\mathbf{H}=0$ lies within the sphere giving $\mathcal{C}\mathpzc{h}=1$, whereas for $H_0 > H_r$ it lies outside the sphere giving $\mathcal{C}\mathpzc{h}=0$.

The nature of the topological and trivial phases can be further revealed by probing their microscopic structure with a conventional adiabatic method. According to Haldane, each phase has its own signature spin texture in momentum space. We again consider spherical surfaces $\mathcal{S}$ and adiabatically ramp the control parameters to their final values on $\mathcal{S}$. The resulting Bloch vectors are then tomographically measured\,\cite{supp}, and ideally point in the same direction as the final $\mathbf{H}$. With a confocal mapping (see\,\cite{supp}), $\mathcal{S}$ can be mapped to the first Brillouin zone (FBZ) of the honeycomb lattice. Therefore, the adiabatically measured ground state vectors on $\mathcal{S}$ can be depicted in the FBZ. Fig.\,3(b) and (c) show the results for two manifolds with \ensuremath{H_{0}/H_{r}=1.2} and 0, corresponding to trivial and topological phases, respectively. By following the orientation of the state-vector along any path starting at \ensuremath{\mathbf{K}} and moving to \ensuremath{\mathbf{K'}} (corners of the FBZ) and back to \ensuremath{\mathbf{K}} one can see that in the topological case the state vector makes one full rotation, while in the trivial case and only tilts away from vertical and then returns, without completing a rotation. These spin texture maps can be used to extract local Berry curvature\,\cite{supp}. As shown in Fig.\,3(c), the resulting  $\mathcal{C}\mathpzc{h}$ from this adiabatic method shows good agreement with the dynamical method of measurement.

Moving beyond the realm of non-interacting systems, we now study the topological phase diagram for an interacting Hamiltonian, obtained by measuring $\mathcal{C}\mathpzc{h}$ in a coupled two-qubit system. The intriguing physics of the topological properties of this kind of interacting system has to date been mostly unexplored, due to experimental challenges. One significant source of challenge is that one needs full control over the entire parameter space, including over any coupling terms in the Hamiltonian. Here we achieve this kind of full control by using a new design for our superconductong qubit, which includes the ability to continuously vary the inter-qubit coupling strength g (we term this new type of qubit the "gmon"\,\cite{Geller2014,Yu2014}).

The Hamiltonian of this system is given by
\begin{equation} \label{2Q}
\mathcal{H}_{2Q}= -\frac{\hbar}{2}[H_0\sigma_1^z + \mathbf{H_1}\cdot\boldsymbol{\sigma_1} + \mathbf{H_2}\cdot\boldsymbol{\sigma_2}-g(\sigma_1^x\sigma_2^x + \sigma_1^y \sigma_2^y)],
\end{equation}
where 1 and 2 refer to qubit 1 (Q1) and qubit 2 (Q2) respectively, and the biasing field $H_0$ is now only applied to Q1.There are equivalent condensed matter systems to which this system can be mapped, as with the Haldane model, as discussed in\,\cite{supp}.However, in the absence of any experimental realization of these models, our experiment is perhaps closer to Feynman's original idea of quantum simulation\,\cite{Feynman1982}, where a controllable quantum system is used to investigate otherwise inaccessible quantum phenomena.

Using the tunable inter-qubit coupling, we can access all regions of the 7-dimensional parameter space of our Hamiltonian. Here we explore spherical manifolds with fixed $(H_{0}, |\mathbf{H_1}|, |\mathbf{H_2}|, g)$, analogous to the single qubit experiment.We perform experiments where both $\mathbf{H_1}=\mathbf{H_2}=\mathbf{H_r}$ are ramped simultaneously with magnitude $|\mathbf{H_r}|=H_{r}$, while $H_0$ and $g$ are zero except during the time $t=0$ to $T_f$, as illustrated in the supplement\,\cite{supp}.  The measured $\mathcal{C}\mathpzc{h}$ is shown in Fig.\,4(b) and (c) for three distinct cuts though this parameter space, as shown by colored rectangles in Fig.\,4(d). For $g=0$ [panel (b)], the two qubits behave independently and the physics is the same as for the single qubit case. Since only Q1 is subject to $H_0$, $\mathcal{C}\mathpzc{h}$ changes
by 1 through the transition $H_0=H_r$.

A new phase with $\mathcal{C}\mathpzc{h}=0$ emerges when the coupling $g$ is large. In Fig.\,4(b) for $g/2\pi = 4\,\textrm{MHz}$, the $\mathcal{C}\mathpzc{h}=0$ phase (blue) is seen at small $H_r$ when $H_r \lesssim g$.  In Fig.\,4(c) this phase also appears when $g \gtrsim H_r$, showing that the transition is interaction-driven and appears when the coupling $g$ becomes dominant.  Because \eqref{2Q} is not SU(2) symmetric, the results do not simply reflect the total spin of the system. However, an intuitive understanding of these phases and transitions can be attained in certain limits: at large $H_r$, the spins align paramagnetically with the field and add up to give $\mathcal{C}\mathpzc{h}=2$. At large $g$, the spins form an entangled singlet which does not respond to the applied field, giving $\mathcal{C}\mathpzc{h}=0$. Away from these limiting cases, these simple arguments are not applicable, but $\mathcal{C}\mathpzc{h}$ remains quantized.

Analytic solutions predicting the phase diagram can be obtained by calculating when points with degenerate ground states cross the spherical manifold\,\cite{supp}. These phase boundaries are depicted in Fig.\,4(d) and show three distinct regions.  As discussed above, the region where $g$ dominates (blue) has $\mathcal{C}\mathpzc{h}=0$, while where $H_r$ dominates (red) $\mathcal{C}\mathpzc{h}=2$.  There is a direct 0 to 2 transition when $H_0=0$, but at finite values the system first goes through the green $\mathcal{C}\mathpzc{h}=1$ region.  This latter behavior is seen in Fig.\,4(c).  The dashed lines in panels (b) and (c) are from this analytic solution, which uses no free parameters, and are in good agreement with the measurements. The deviations are mainly systematic errors, due to crosstalk between simultaneous control pulses. As shown in Fig.\,4(a), the points of ground state degeneracy are located on the $z$-axis of the $\mathbf{H_r}$-space\,\cite{supp}. Sub-figures A, B and C correspond to the dots on Fig.\,4(c), where $g$ is small. In this limit, $H_0$ affects the energy of only one qubit, and increasing it moves only one monopole past the surface (C). For D, E, and F where instead $H_0$ is small, increasing $g$ furthers the monopole separation, eventually moving both monopoles outside the surface (F).

An important benefit of working with a fully controllable Hamiltonian, as here, is that a number of different condensed matter systems can be mapped onto this model system. For our 2-qubit system, we show\,\cite{supp} that the system can be mapped to either an interacting model, or alternatively a 4-band non-interacting electron model that is a non-trivial extension of the two-band Haldane model. In general, with \(n\) qubits one can study topological phases in non-interacting \(2^n\)-band models, an otherwise daunting experimental task. Perhaps more interesting will be to use qubit systems to study the topological phases of interacting spin-\(1/2\) systems, where tantalizing evidence for fractionalization has been found\,\cite{Anatoli}.

\textbf{Acknowledgments:} We acknowledge fruitful discussions with R. Lutchyn, J. Moore, C. Nayak, M. Niu, A. Rahmani, T. Souza, M. Vavilov, D. Weld, and A. Yazdani. This work was supported by the NSF under grants DMR-0907039 and DMR-1029764, the AFOSR under FA9550-10-1- 0110, and the ODNI, IARPA, through ARO grant W911NF-10-1-0334. Devices were made at the UCSB Nanofab Facility, part of the NSF-funded NNIN, and the NanoStructures Cleanroom Facility.


\begin{thebibliography}{30}
\bibitem [{ch3()}]{ch3D}%
  \BibitemOpen
  \href@noop {} {}\bibinfo {note} {This form of the Gauss's law in which the
  Berry curvature is treated as a 3-vector is only valid in a three-dimensional
  parameter space. Generalization of this logic to higher-dimensional manifolds
  remains work in progress.}\BibitemShut {Stop}%
\end{thebibliography}
\end{document}


\title{Supplementary materials:
Observation of topological transitions in interacting quantum circuits}

\author{P. Roushan$^{1}$}
\thanks{These authors contributed equally to this work.}
\author{C. Neill$^{1}$}
\thanks{These authors contributed equally to this work.}
\author{Y. Chen$^{1}$}
\thanks{These authors contributed equally to this work.}
\author{M. Kolodrubetz$^{2}$}
\author{C. Quintana$^{1}$}
\author{N. Leung$^{1}$}
\author{M. Fang$^{1}$}

\author{R. Barends$^{1}$}
\author{B. Campbell$^{1}$}
\author{Z. Chen$^{1}$}
\author{B. Chiaro$^{1}$}
\author{A. Dunsworth$^{1}$}
\author{E. Jeffrey$^{1}$}
\author{J. Kelly$^{1}$}
\author{A. Megrant$^{1}$}
\author{J. Mutus$^{1}$}
\author{P. O'Malley$^{1}$}
\author{D. Sank$^{1}$}
\author{A. Vainsencher$^{1}$}
\author{J. Wenner$^{1}$}
\author{T. White$^{1}$}

\author{A. Polkovnikov$^{2}$}
\author{A. N. Cleland$^{1}$}
\author{John M. Martinis$^{1}$}

\email{martinis@physics.ucsb.edu}

\affiliation{$^{1}$Department of Physics, University of California, Santa Barbara,
CA 93106-9530, USA}

\affiliation{$^{1}$Department of Physics, Boston University, Boston,
MA 02215, USA}
\maketitle
\textbf{}
\tableofcontents

\clearpage
\section{\textcolor{SECTIONCOLOR}{1. The gmon qubits}}

\begin{figure*}
\begin{centering}
\includegraphics[width=178mm]{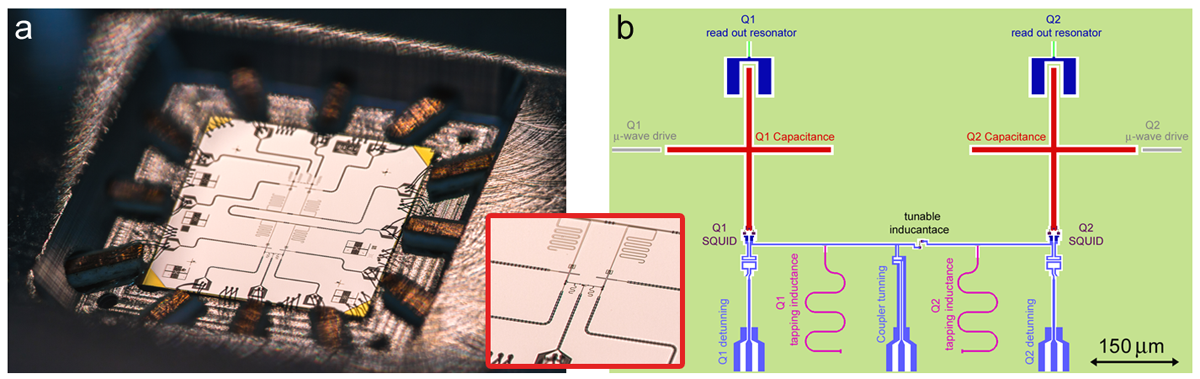}
\par\end{centering}
\caption{
\textbf{Device architecture.} \textbf{(a)} Optical image of the device showing two coupled gmon transmons on the top half of the chip and the two coupled gmons used in this work on the lower half (zoomed-in view in inset). \textbf{(b)} The layout of the two-qubit gmon system. We supply bias currents using the lower blue lines to tune the inductance of the coupler junction (middle) and the qubit frequencies (left, right). We apply microwave pulses to each qubit via the gray trace. We read out the state of the qubits dispersively via readout resonators: each qubit is capacitively coupled to a resonator (green lines; meandered lines in inset of (a)). }
\label{fig:concept}
\end{figure*}

\subsection{\textcolor{SUBSECTIONCOLOR}{1.1. The gmon coupling architecture}}

In this work we implemented an adjustable inductive coupling between two qubits. Adjustable coupling has typically been difficult with superconducting qubits, as fixed capacitive coupling may only be modified by detuning, so it has the problems of limited on/off range and crosstalk.  Here we use a novel qubit design called the gmon, which allows a continuous variation of the inter-qubit coupling strength $g$ over nanosecond time scales without any degradation in the coherence of the constituent Xmon qubits\,\cite{Geller2014,Yu2014}. The adjustable inductive coupling between the transmons allows $g/2\pi$ to be varied between $-5$ MHz and $55$ MHz, including zero, without changing the bare qubit frequencies. The device was fabricated using standard optical and e-beam lithography techniques, discussed in recent works of our group such as\,\cite{RamiNature2014}. The experiment was performed at the base temperature of a dilution fridge ($\sim$20 mK).

\subsection{\textcolor{SUBSECTIONCOLOR}{1.2. Basic design principle of the gmon}}

As shown in Fig. S1, the gmon design is based on the Xmon qubit design. One important feature of the Xmon design\,\cite{RamiPRL} is the single-ended ground in contrast to differential or floating grounds. In the absence of adjustable coupling, the SQUID loops (Fig. S1(b)) would be directly connected to the ground plane. This design feature gives us the ability to capacitively couple qubits with elements such as the drive lines, the readout resonators, and nearby qubits. In the gmon architecture, instead of immediately terminating the qubit SQUID to ground, we add a linear inductor (the meandering CPW element colored in purple and labeled "tapping inductance") between the SQUID and ground(CPW stands for coplanar waveguides). This creates a node (where the purple CPW meets the horizontal blue CPW) that allows us to couple the two qubits. The two qubits then can be connected with a CPW line. This connecting line is interrupted with a Josephson junction, which acts as a tunable inductor that can be used to tune the inter-qubit coupling strength \(g\), hence the name gmon.

The basic operation of the gmon can be understood from a simple linear circuit model. An excitation created in Q1 will mostly flow to ground through its tapping inductance, but a small fraction will flow to the tapping inductance of Q2, generating a flux in Q2. The mutual inductance resulting from the flux in Q2 due to an excitation current in Q1 can be calculated from simple current division, and the coupling strength is proportional to this mutual inductance to high accuracy\,\cite{Geller2014,Yu2014}. The current division ratio, which sets the coupling strength \(g\), can be varied by changing the superconducting phase difference across the tunable inductance. This is done by flux biasing its junction, using the current line labeled "coupler tuning" in panel (b).

An important advantage of this architecture is that it prevents crosstalk, a serious hurdle for many other experimental works. Coupling the qubits at the nodes between the SQUID and the tapping inductance allows a DC current to set the coupling strength. Because of the open loop of each qubit (due to capacitance) this DC current cannot flow to the SQUID and change the qubit frequencies. Thus, the two capacitors act as DC blocks. This key ingredient of the gmon design minimizes the crosstalk between the qubits and the coupler.

\subsection{\textcolor{SUBSECTIONCOLOR}{1.3. Coherence of the gmon}}

One major concern of the coupler circuit is degradation of the qubit coherence. The gmon architecture required adding CPW lines to connect qubits, ground plane cross-overs which involve dielectrics, and a tunable inductance. If the capacitive loss due to these elements are not properly considered, the coherence of the system could be substantially degraded. With the gmon design, the fraction of the qubit energy stored in these coupling elements scales as the square of the ratio of the tapping inductance to the qubit SQUID inductance, which is \(1/2000\). Therefore these elements do not affect the qubit coherence\,\cite{Yu2014}. Furthermore, to avoid inductive loss, we used a relatively small mutual coupling of around 1\,pH to the coupler tuning line. This coupling places an upper bound of 200\,\(\mu\)s on the energy relaxation time \(T_1\). The average measured \(T_1\) for our device was around 10\,\(\mu\)s, independent of the coupling strength. This is comparable to the performance of Xmon qubits with the same geometry and material. As demonstrated with Xmon qubits\,\cite{RamiNature2014}, the coherence can be improved by widening the capacitor and using MBE-grown aluminum films.

\vspace{25mm}
\section{\textcolor{SECTIONCOLOR}{2. Mapping the Single-Qubit Hamiltonian to the Haldane Model and Adiabatic Measurement of the Chern number}}

\subsection{\textcolor{SUBSECTIONCOLOR}{2.1. Haldane model}}

To show that the quantum Hall effect could be achieved without a global magnetic field, Haldane introduced a non-interacting Hamiltonian\cite{Haldane1988}, which served as the cornerstone of future topological band studies. He introduced a massive Dirac Hamiltonian with different mass terms at the two non-equivalent corners of the Brillouin zone \(\mathbf{K},\mathbf{K'}\). Near these points the Hamiltonian is given by
\begin{equation}
\mathcal{H}_{G}^{\pm}(k_{x}^{\pm},k_{y}^{\pm})=\hbar v_{F}(k_{x}^{\pm}\sigma^{x}\pm k_{y}^{\pm}\sigma^{y})+(m_{0}\mp m_{t})\sigma^{z},\label{graphene}
\end{equation}
where \ensuremath{v_{F}} is Fermi velocity, and \ensuremath{k_{x}^{+}(k_{x}^{-}),k_{y}^{+}(k_{y}^{-})} are measured from two non-equivalent corners of the Brillion zone \ensuremath{\mathbf{K}(\mathbf{K'})}. \ensuremath{m_{0}} is the mass associated with inversion symmetry breaking, and \ensuremath{m_{t}}  corresponds to a second-neighbor hopping in a local magnetic field. The key prediction of the Haldane model is that if \(m_0/m_t>1\)  the system is in a trivial insulating phase, and otherwise in a topological phase, where edge states and quantized conductance appear.

Using a confocal mapping, discussed below, one can recast Eq. \eqref{graphene} into the single qubit Hamiltonian of the main text (eqn. (3)). For convenience we re-parameterize that equation in terms of a field \(H_0\) along the \(z\)-axis, and a radial field \(H_r\) with orientation given by \(\theta,\phi\), such that \(\mathbf{H}=(H_r \sin \theta \cos \phi,H_r \sin \theta \sin \phi,H_0-H_r \cos \theta)\). Then, for \(\theta\) values close to 0 and $\pi$ the single qubit Hamiltonian becomes

\begin{equation}
\mathcal{H}_{S}^{\pm}(H_{0},H_{r},\theta,\phi)=-\frac{\hbar}{2}(H_{r}(\sin(\theta)\cos(\phi)\sigma^{x}+\sin(\theta)\sin(\phi)\sigma^{y})+(H_{0}{\pm}H_{r})\sigma^{z}).\label{sm}
\end{equation}

 By comparing \ensuremath{H_{S}^{+}(H_{S}^{-})} to \ensuremath{H_{G}^{+}(H_{G}^{-})}, it becomes evident that \ensuremath{H_{0}/H_{r}} in the qubit system plays the same role as \ensuremath{m_{0}/m_{t}} in the Haldane model. The fact that the topological phase transition occurs at \ensuremath{H_{0}/H_{r}}=1 is consistent with the Haldane model, where the transition takes place at \ensuremath{m_{0}/m_{t}}=1. Similar to the Haldane model, where \(k_x\) and \(k_y\) span a manifold of states in the Brillouin zone,  \(\theta\) and \(\phi\) span a manifold in the parameter space of the qubit system. With this mapping, the two distinct phases observed in Fig. 3(a) of the main text correspond to the topological and trivial phases in the Haldane model.

\begin{figure*}
\begin{centering}
\includegraphics[width=178mm]{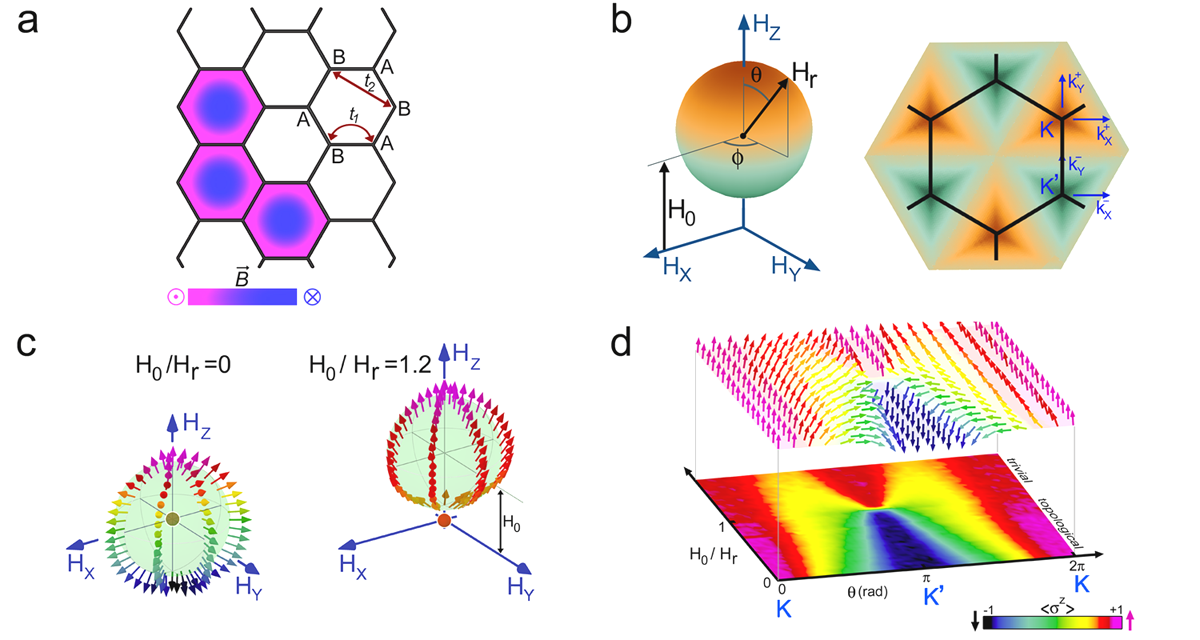}
\par\end{centering}
\caption{
\textbf{Experimental visualization of the topological phases and their evolution across the transition.} \textbf{(a)} In the Haldane model of graphene, in addition to the nearest neighbor hopping ($t_1$), a second neighbor hopping ($t_2$) is also considered with a variable phase $\phi$ controlled by the locally-varying flux through the plaquette, as well as a sublattice "mass" $m_0$ corresponding to a difference in chemical potentials between the sublattices. This system is topologically trivial if $|m_0| > |m_t=3 \sqrt 3 t_2 \sin \phi|$ and non-trivial otherwise. \textbf{(b)} A color-assisted representation of the mapping from a sphere parameterized by \(H_0,H_r,\theta,\phi\) to the hexagonal Brillouin zone of graphene. \textbf{b.} With adiabatic state preparation, the state of the qubit was prepared and measured over a grid on the surface of the parameter space spheres. Selected adiabatic Bloch sphere vectors are shown for \(H_0/H_r=0\) and 1.2. \textbf{(d)} With adiabatic state preparation, the state of the qubit was prepared and measured along the \(\phi=0\)\,meridian for various \(H_0/H_r\) values. The Bloch sphere states are presented with arrows whose colors indicate their \(\left<\sigma^z\right>\) values. The topological and trivial phase each has its own signature textures. By following the orientation of the state-vector along any path starting from \ensuremath{\mathbf{K}} to \ensuremath{\mathbf{K'}} and back to \ensuremath{\mathbf{K}} one can see that in the topological case the state-vector fully winds around; however, for the trivial phase it only tilts away from the north pole of the Bloch sphere and comes back without winding around.}
\label{fig:concept}
\end{figure*}

\vspace{5mm}

\subsection{\textcolor{SUBSECTIONCOLOR}{2.2. Adiabatic measurement, confocal mapping, and direct measurement of the Chern number}}

To visualize how the qubit and Haldane model are topologically related, we now explicitly construct a mapping between the single qubit parameter space and momentum space in the Haldane model. We use this to map the qubit Bloch vector measured by adiabatic state preparation to the first Brillouin zone of graphene and then compute its Chern number, thus completing the analogy with the Haldane model. By using a confocal mapping, the parameter space points can be mapped to the hexagonal Brillouin zone of the honeycomb lattice of graphene (from \textit{H}-space to k-space). The mapping places the points on the northern part of the spherical manifold of radius unity around the \ensuremath{\mathbf{K}} point of the Brillouin zone and the southern hemisphere points around the \ensuremath{\mathbf{K'}} point.  For the northern hemisphere

\begin{eqnarray}
\rho &=& r(\phi) \tan \frac{\theta}{2} \nonumber\\
\varphi &=& \phi
\end{eqnarray}
where \(\theta\) and \(\phi\) are the spherical coordinates of the northern hemisphere of the sphere in the parameter space, \(\rho\) and \(\varphi\) are the polar coordinates of the mapped circle, and \(r\)  is given by
\begin{eqnarray}
r(\phi) &=& \frac{b \sin(\pi/6)}{\sin (5\pi/6 - \phi)} \text{for } 0\leq \phi < 2\pi/3 \nonumber\\
&=& \frac{b \sin(\pi/6)}{\cos \phi} \text{ for } 2\pi/3 \leq \phi < 4\pi/3  \nonumber\\
&=& \frac{b\sin(\pi/6)}{\sin (\phi - 7\pi/6)} \text{ for } 4\pi/3 \leq \phi < 2\pi,
\end{eqnarray}
where \(b=|\mathbf{K}-\mathbf{K'}|\). The mapping of the southern hemisphere takes a similar form. This mapping is illustrated in Fig. S2(b), and covers only one third of the first Brillouin zone  (FBZ). To cover the entire FBZ, the mapping was repeated three times. As the colors in Fig. S2(b) show, the north pole in the parameter space maps to three equivalent \ensuremath{\mathbf{K}} points at the corners of FBZ and the south pole to the three \ensuremath{\mathbf{K'}} points. With the mapping principle explained, now we can "move" the ground states that are measured adiabatically on the spherical surfaces in the parameter space and place them in the FBZ of the honeycomb lattice. Fig. S2(c) and (d) show the results for \ensuremath{H_{0}/H_{r}=1.2}, and 0, corresponding to trivial and topological phases, respectively.

Knowing the ground state of the system for each \(k_x\) and \(k_y\) point in the FBZ, $\mathcal{C}\mathpzc{h}$ can be calculated directly from
\begin{equation}\label{chern2}
\mathcal{C}\mathpzc{h} = \frac{1}{4\pi} \int_{B.Z} \boldsymbol{\sigma}(\mathbf{k}) \cdot \left[ \frac{\partial\boldsymbol{\sigma}(\mathbf{k})}{\partial k_x} \times \frac{\partial \boldsymbol{\sigma}(\mathbf{k})}{\partial k_y} \right] d^2 \mathbf{k}.
\end{equation}
Using this relation $\mathcal{C}\mathpzc{h}$ numbers shown in Fig. 3(d) of the main text as well as the ones on panels (c) and (d) of Fig. S2 are calculated. While without any mapping, $\mathcal{C}\mathpzc{h}$ could be calculated by modifying \eqref{chern2} to make it appropriate for spherical coordinates, mapping from the sphere to a 2D plane allowed us to use \eqref{chern2} directly. It is interesting to note that due to the topological nature of the phases, the details of the mapping do not matter, and other mappings could have worked as well.

\vspace{5mm}
\subsection{\textcolor{SUBSECTIONCOLOR}{2.3. Discussion}}

We note that the Haldane model consists of a half filled lattice of non-interacting spins, while we constructed a manifold of ground states by measuring the qubit over a closed surface. This difference is resolved by considering a mapping of the ground state manifold to the valence band of graphene, while the excited state manifold maps to the conduction band. Therefore, by probing the entire parameter manifold of the qubit ground state, we are probing the entire valence band. This distinction is why $\mathcal{C}\mathpzc{h}$ of the electronic system can be measured with a single measurement, since all of the electronic states at different momenta are filled and hence probed simultaneously. On the other hand, the qubit can be measured only at a single point of the qubit system's parameter space at a given time, which is why all the parameter angles must be probed separately and integrated to give $\mathcal{C}\mathpzc{h}$.

\clearpage
\section{\textcolor{SECTIONCOLOR}{3. Loci of Monopole Singularities for the Two-Qubit System}}

We have used the electromangetic analogy extensively in the main text to plot the (theoretical) locations of the magnetic monopoles in parameter space; we now show how one can identify their locations.  We begin by pointing out that the magnetic monopole density can also be written as $\rho_m = \frac{1}{2\pi} \nabla \cdot (\nabla \times {\bf A})$, a seeming contradiction given that the divergence of a curl is known to be zero!  The resolution to this seeming contradiction is that $\nabla \cdot (\nabla \times {\bf A})$ is indeed zero whenever the function ${\bf A}$ is smooth.  However, near ground state degeneracies one cannot pick a smooth choice of gauge (even locally), since the ground state undergoes a sharp change.  Therefore $\rho_m$ is allowed to be non-zero if and only if the ground state is degenerate.  Thus, one can reduce the problem of finding the magnetic charge density to the simpler one of locating the ground state degeneracies.

For general two qubit Hamiltonians, degeneracies can be readily located numerically using conventional techniques to minimize the ground state energy gap.  However, for the specific case of our cylindrically symmetric two-qubit Hamiltonian, we can solve the problem analytically.  As a reminder, the Hamiltonian of interest is
\begin{equation}
\mathcal{H} = -\frac{1}{2} \left[ H_r \hat n (\theta,\phi) \cdot ({\bf \sigma_1} + {\bf \sigma_2}) + H_0 \sigma^z_1 + g(\sigma^x_1 \sigma^x_2 + \sigma^y_1 \sigma^y_2) \right] ~.
\label{eq:H_2Q_app}
\end{equation}
This Hamiltonian has $U(1)$ invariance, meaning that the Hamiltonian at $\phi=0$ can be mapped to arbitrary $\phi$ using
\begin{equation}
\mathcal{H}(\theta,\phi)=e^{i \phi \sigma^z_\mathrm{tot} / 2} H(\theta,0)e^{-i \phi \sigma^z_\mathrm{tot} / 2} ~,
\end{equation}
where $\sigma^z_\mathrm{tot}=\sigma^z_1+\sigma^z_2$. While this invariance is computationally useful, it does not lead to any additional conservation laws, so on general grounds one does not expect to find degeneracies of our Hamiltonian (a $4\times 4$ matrix) in the absence of symmetry. However, at $\theta=0$ and $\pi$, the $U(1)$ invariance becomes a $U(1)$ symmetry, $\sigma^z_\mathrm{tot}$ is a conserved quantity, and this enables ground state degeneracies.

The values $\theta=0$ and $\pi$ lie along the $z$-axis, so we reparameterize the Hamiltonian along this axis as
\begin{equation}
\mathcal{H} = -\frac{1}{2} \left[ H_z (\sigma_1^z + \sigma_2^z) + H_0 \sigma^z_1 + g(\sigma^x_1 \sigma^x_2 + \sigma^y_1 \sigma^y_2) \right] ~.
\label{eq:H_2Q_z}
\end{equation}
Since total spin along the $z$ axis is now conserved, there are two obvious eigenstates: 
with energies $E_{\up \up/\down \down} = \pm (H_z + H_0/2)$.
Within the $s^z_\mathrm{tot}=0$ sector, the Hamiltonian reduces to

\begin{equation}
\mathcal{H}_{\up\down} = -\frac{1}{2} \left( \begin{array}{cc} H_0 & 2g \\ 2g & -H_0 \end{array}\right)~,
\end{equation}
which has eigenenergies $E_{\up \down} = \pm \sqrt{H_0^2/4 + g^2}$.

The ground state energy levels of these two sectors are degenerate when $|H_z + H_0 / 2| = \sqrt{H_0^2/4 + g^2}$, from which we find
\begin{equation}
H_z^\mathrm{deg} = \frac{-H_0 \pm \sqrt{H_0^2 + 4 g^2}}{2} ~.
\end{equation}
Having located these degeneracies, we can identify their magnetic monopole charges as $Q_m = 1$ based on the jump $\mathcal{C}\mathpzc{h}$ that we find (experimentally and theoretically) at the topological transitions.

This $U(1)$ symmetry at the poles was useful for our analysis, but it is not likely to exist for more complicated cases in which the Hamiltonians are not quite so exquisitely tunable.  Therefore, by our above logic, we might argue that the degeneracies should go away if there are no longer any symmetries protecting them.  However, our topological properties are robust against any perturbation, so despite the loss of symmetry, the degeneracies may drift around, but they do not disappear!  This is a situation in which the degeneracies are protected not by a global symmetry, but rather by an emergent topological protection
\cite{LeekScience}. Breaking the $U(1)$ invariance of the model -- for example by adding a $\sigma^x_1$ term to the Hamiltonian -- would disrupt the measurement.  In our case, such a symmetry-breaking term does not present any fundamental challenges.  It would simply add $\phi$ dependence to the Berry curvature, now requiring data to be taken for ramps of $\theta$ at multiple values of $\phi$ to allow integration over this direction as well.  While it is certainly more time consuming to take this extra data, fundamentally it is no more difficult.

\clearpage
\section{\textcolor{SECTIONCOLOR}{4. Hitchhiker's Guide to the Chern number and Berry Curvature}}

In the main paper, we defined $\mathcal{C}\mathpzc{h}$ in terms of something called the ``Berry curvature,'' which may have seemed mysterious. Here, we introduce the concept of Berry curvature in the context of the more familiar Berry phase studied in adiabatic quantum mechanics. This will allow us to understand both the geometric and ``electromagnetic'' interpretations of Berry physics in an intuitive but quantitative way, which will in turn lead to the topological interpretation used in the main paper of $\mathcal{C}\mathpzc{h}$ as a count of the number of degeneracies enclosed by a ground state manifold. We note that our pedagogical treatment of degeneracies as sources of a curvature field largely follows the original exposition of the Berry phase and curvature by M. V. Berry \cite{Berry1984}.

\noindent
\subsection{\textcolor{SUBSECTIONCOLOR}{4.1 Berry connection, phase, curvature and all that}}

\subsubsection{\textcolor{SUBSUBSECTIONCOLOR}{4.1.1 Berry connection and phase}}
Suppose we have a Hamiltonian that depends on a set of external parameters, which we describe by the parameter space vector ${\bf{R}}$, with corresponding ground states $\ket{\psi_0(\mathbf R)}$; i.e., $\mathcal{H}(\mathbf{R})\ket{\psi_0(\mathbf{R})} = E_0 \ket{\psi_0(\mathbf{R})}$. An example would be the three-dimensional (3D) parameter space associated with a single-qubit Hamiltonian in a rotating frame, $\mathcal{H}(\mathbf{R}) = -\frac{\hbar}{2}(H_X \sigma^x + H_Y \sigma^y + H_Z \sigma^z)$, with $\mathbf{R} = (H_X,H_Y,H_Z)$. Alternatively, if we take the rotating field in spherical coordinates, the natural parameters are magnitude $H_r \equiv |\mathbf{H_r}|$ and angles $\theta$ and $\phi$ (as in the main text). The Berry connection (from which the Berry curvature is defined) associated with the ground state manifold is then
\begin{equation}
{\bf A} = i \braxket{\psi_0}{\nabla_{\bf R}}{\psi_0},\label{eq:berry_connection}
\end{equation}
which when integrated around a closed path $C$ in parameter space yields the celebrated geometric Berry phase associated with that path \cite{Berry1984,Griffiths}
\begin{equation}
\gamma(C) = \oint_C \mathbf{A} \cdot d\mathbf{R}.\label{berry_phase_integral}
\end{equation}
This fact can be derived from the Schr\"{o}dinger evolution of a quantum state as $C$ is traversed in parameter space in the adiabatic limit, and is independent of whatever dynamical phase is accumulated throughout the closed trajectory. However, it is not necessary to understand the phenomena of Berry phase from the perspective of the time-evolution of adiabatic systems -- one can simply view it as a consequence of the geometry of an eigenstate manifold, which will soon become apparent in our discussion.

\subsubsection{\textcolor{SUBSUBSECTIONCOLOR}{4.1.2 Geometric interpretation of Berry connection}}
The Berry connection $\mathbf{A}$ is an interesting construct because the meaning of the expression $\nabla_\mathbf{R}\ket{\psi_0}$ is ambiguous when only $\mathcal{H}(\mathbf{R})$ is given: unlike the coordinates $\mathbf{X}$ of real space, where a state $\ket{\psi}$ can be expanded as a wavefunction of spatial coordinates and $\nabla_\mathbf{X}$ is a natural operator on these wavefunctions, here it is instead the Hamiltonian itself that is a function of the parameter space coordinates $\mathbf{R}$. A manifold of ground states can be associated with a manifold in parameter space via the defining eigenvalue condition $\mathcal{H}(\mathbf{R}) \ket{\psi_0(\mathbf{R})} = E_0(\mathbf{R})\ket{\psi_0(\mathbf{\mathbf{R}})}$; however, although the states $\ket{\psi_0(\mathbf{R})}$ all live in the same Hilbert space, this eigenvalue condition does not tell us the phase of $\ket{\psi_0(\mathbf{R})}$ at different $\mathbf{R}$. In other words, we must specify what is essentially a choice of gauge when it comes to relative phase relations, and since this choice can be made arbitrarily, we cannot expect it to have any intrinsic physical meaning. Once a choice is made for these phases however, the name ``connection'' for $\mathbf{A}$ signifies that $\mathbf{A}$ encodes a way to equate (or ``connect'') ground state vectors at two nearby points $\mathbf{R}$ and $\mathbf{R} + d\mathbf{R}$ in parameter space, analogous to the differential geometric notion of parallel transport of tangent vectors along a manifold \footnote{In the language of geometry, $A_\mu$ intuitively gives (up to a factor of $i$) the component of $\ket{\psi_0}$ that one must subtract from $\partial_\mu \ket{\psi_0}$ in order to turn the derivative $\partial_\mu\ket{\psi_0}$ into a ``covariant'' derivative $D_\mu\ket{\psi_0}$ with the property that $\ket{D_\mu{\psi_0}}$ is orthogonal to $\ket{\psi_0}$, which defines what it means to keep a quantum state ``parallel'' as $\mathbf{R}$ is moved from $\mathbf{R}$ to $\mathbf{R} + d\mathbf{R}$.}.

\subsubsection{\textcolor{SUBSUBSECTIONCOLOR}{4.1.3 Berry connection as a vector potential: deriving an observable field}}
In light of its dependence on $\nabla_\mathbf{R}$, ${\bf A}$ is therefore \emph{gauge-dependent}. The remarkable fact however, as realized by Berry, is that its integral around a loop is actually gauge-\emph{independent} (modulo $2\pi$), and can therefore be measured. This is easily seen: suppose we change our definition of $\ket{\psi_0(\mathbf{R})}$ by an arbitrary local phase factor, $\ket{\psi_0(\mathbf{R})} \to e^{i\Gamma(\mathbf{R})}\ket{\psi_0(\mathbf{R})}$. Then by equation (\ref{eq:berry_connection}), $\mathbf{A}(\mathbf{R})$ is modified by the addition of the term $- \nabla_\mathbf{R} \Gamma(\mathbf{R})$, which integrates to zero around a closed path. The observant reader will notice that this takes the same form as the change of the magnetic vector potential under a gauge transformation. Recalling that the magnetic field is a gauge-invariant (i.e., directly measurable) quantity derivable from the magnetic vector potential, this motivates us to follow our experience with classical electromagnetism and define the analogue of the magnetic field, $\mathbf{B} \equiv \nabla \times \mathbf{A}$. This will allow us to rewrite the integral (\ref{berry_phase_integral}) defining the Berry phase in terms of an \emph{observable} integrand $\mathbf{B}$. We will see that this ``Berry field'' has the interpretation of intrinsic curvature of the ground state manifold. In addition, this endeavor will expose some interesting physics, including the main topic of our work: topological transitions (jumps in $\mathcal{C}\mathpzc{h}$) associated with degeneracy points.

Continuing the analogy, where for simplicity we consider a 3D parameter space, we obtain the Berry field from the Berry connection:
\begin{equation}
\mathbf{B}(\mathbf{R}) \equiv \nabla_\mathbf{R}\times \mathbf{A}(\mathbf{R}).\label{eq:berry_curvature}
\end{equation}
The Berry curvature field $\mathbf{B}$ is the vector form of what is known as the Berry curvature tensor, defined for general dimensionality and coordinate parametrizations by the antisymmetric tensor $\mathcal{B} \equiv \partial_\mu A_\nu - \partial_\nu A_\mu$ generalizing the curl:
\begin{equation}
\mathcal{B} = \left( \begin{array}{ccc}
\mathcal{B}_{xx} & \mathcal{B}_{xy} & \mathcal{B}_{xz} \\ \mathcal{B}_{yx} & \mathcal{B}_{yy} & \mathcal{B}_{yz} \\ \mathcal{B}_{zx} & \mathcal{B}_{zy} & \mathcal{B}_{zz}
\end{array} \right) = \left( \begin{array}{ccc}
0 & B_{z} & -B_{y} \\ -B_{z} & 0 & B_{x} \\ B_{y} & -B_{x} & 0
\end{array} \right) ~;
\end{equation}
that is, $\mathbf{B} = (B_x, B_y, B_z) = (\mathcal{B}_{yz},\mathcal{B}_{zx},\mathcal{B}_{xy})$. In our case, for short we will simply call $\mathbf{B}$ the Berry curvature.
\subsubsection{\textcolor{SUBSUBSECTIONCOLOR}{4.1.4 Geometric interpretation of the Berry curvature field}}
The Berry phase associated with a closed path can now be calculated from (\ref{berry_phase_integral}) using Stokes' theorem by integrating the Berry curvature over a bounding surface,
\begin{equation}
\gamma(C) = \iint_\mathcal{S} \mathbf{B}(\mathbf{R})\cdot d\mathbf{S},\label{berry_phase_surface}
\end{equation}
where $\mathcal{S}$ is a surface manifold in parameter space whose boundary is $C$. This is the direct analogue of a charged particle acquiring an Aharanov-Bohm phase when its path encloses a magnetic flux. However, the Berry curvature is a local geometric property, and for 2D manifolds can be physically measured through equation (2) of the main text. Intuitively, the Berry curvature at $\mathbf{R}$ is equal to the ratio of the geometric phase accumulated over a loop surrounding $\mathbf{R}$ to the parameter space area enclosed by that loop, in the limit that the size of the loop goes to zero; in other words, it locally measures the noncommutativity of parallel transport, which manifests itself as a local ``twisting and turning'' of the state vector in parameter space via the accumulation of Berry phase. This is analagous to the fact that carrying a tangent vector on a geodesic triangle on the surface of a sphere causes the tangent vector to change direction when the triangular path returns to its starting point, even though locally the vector is always transported in a parallel fashion. The analogue of the Lorentz force [equation (2) of the main text] for the ``magnetic field'' $\mathbf{B}$ is related to this geometry-induced ``deflection.''

\subsubsection{\textcolor{SUBSUBSECTIONCOLOR}{4.1.5 From local to global properties: the Chern number $\mathcal{C}\mathpzc{h}$}}
One of the main points of this work is that through the analogy to electromagnetism, we can understand how to relate these geometric properties to topological properties of the ground state manifold as a whole. The natural question is then what generates the field $\mathbf{B}$ -- is it the current of ``charged particles,'' or an analogue to the magnetic monopole? Consideration of this question leads us to a Gauss's law interpretation of $\mathcal{C}\mathpzc{h}$, whose definition we repeat here [equation (1) of the main text]:
\begin{equation}
\mathcal{C}\mathpzc{h}(\{\ket{\psi_0}\}) \equiv \frac{1}{2\pi} \oint_{\mathcal{S}}  {\bf B} \cdot d\mathbf{S}.
\end{equation}
This is an integral of $\mathbf{B}$ over a closed (meaning no boundary curve $C$) ground state manifold in parameter space, and gives \emph{nonlocal} information about this manifold in the form of a discrete integer through the Chern theorem \cite{Bernevig}. To deduce the quantization of $\mathcal{C}\mathpzc{h}$, we will use an argument similar to Dirac's argument \cite{Dirac1930} showing that the magnetic monopole charge is quantized. After that, we will explicitly relate this quantized value to the number of enclosed ``magnetic monopoles'' in parameter space. As usual, we restrict ourselves to 2D surfaces in a 3D parameter space.

The astute reader may wonder, given the definition of $\mathbf{B} = \nabla \times \mathbf{A}$, why $\mathcal{C}\mathpzc{h}$ is not simply zero -- after all, a simple application of Stokes' theorem shows that the integral of the curl of a function over any closed surface must vanish: imagine forming an arbitrary closed path $C$ on the surface manifold $\mathcal{S}$, and let $\mathcal{S}_1$ and $\mathcal{S}_2$ be the two surfaces into which $C$ divides $\mathcal{S}$. Taking into account the relative orientation of the two surfaces we then have
\begin{equation}
\mathcal{C}\mathpzc{h} = \frac{1}{2\pi}\left(\int_{\mathcal{S}_1} \mathbf{B}\cdot d\mathbf{S} - \int_{\mathcal{S}_2} \mathbf{B}\cdot d\mathbf{S} \right).\label{Ch_is_integer}
\end{equation}
A naive application of Stokes' theorem would say that each term is equal to the line integral of $\mathbf{A}$ around the same path, but with opposite signs, leading to $\mathcal{C}\mathpzc{h} = 0$. However, this assumes that a single Berry connection (i.e., vector potential) can be defined over the entire manifold with some sufficient smoothness condition. Since Stokes' theorem can be intuitively understood by dividing the surface of integration into infinitely many infinitesimal circulation integrals of $\mathbf{A}$ and noting that neighboring circulations cancel everywhere except along the surface boundary $C$, if there is a singularity in $\mathbf{A}$ then Stokes' theorem will break down. It then becomes a topological constraint on any vector potential covering $\mathcal{S}$ that there must be a singularity in $\mathbf{A}$ somewhere on the surface, which allows for the possibility of non-zero $\mathcal{C}\mathpzc{h}$. The interesting fact is that the location of this singularity depends on the choice of vector potential (i.e., is gauge-dependent), but its existence does not depend on the choice of gauge. We note that a similar argument with what is now known as the Aharanov-Bohm phase associated with a physical magnetic field leads to Dirac's quantization condition for real magnetic monopoles \cite{Dirac1930}.

However, there is still a constraint on the possible values of $\mathcal{C}\mathpzc{h}$. Looking again at equation (\ref{Ch_is_integer}), since the geometric phase (\ref{berry_phase_integral}) accumulated by traversing $C$ is physically observable (modulo $2\pi$), using Stokes' theorem for each surface with its own vector potential it must be the case that the flux of $\mathbf{B}$ through $\mathcal{S}_1$ differs from the flux of $\mathbf{B}$ through $\mathcal{S}_2$ by a factor of $2\pi N$, where $N = \mathcal{C}\mathpzc{h}$ is an integer.

$\mathcal{C}\mathpzc{h}$, which is a property of the entire ground state manifold and cannot be probed locally, is therefore an example of a discrete topological invariant. In particular, $\mathcal{C}\mathpzc{h}$ is robust to perturbations to the parameter space manifold, and it is reasonable to expect that it can only undergo transitions between different quantized values when there is singular behavior on the surface $\mathcal{S}$. 
In the next section, we will see that in our experiment, these singularities are precisely the locations of ground state degeneracy in the Hamiltonian, and will show that when the degeneracies considered are two-fold, $N$ is in fact precisely equal to the number of two-fold degeneracies enclosed by the surface.

\noindent

\subsection{\textcolor{SUBSECTIONCOLOR}{4.2 Topological interpretation of $\mathcal{C}\mathpzc{h}$ in terms of enclosed degeneracies}}

What determines this mysterious integer $N$, and how can we observe it? The concept of $\mathcal{C}\mathpzc{h}$ as a topological invariant is reminiscent of the Gauss-Bonnet theorem from differential geometry, which relates the integral of the Gaussian curvature over a closed surface to its topological genus. In the case of the Gauss-Bonnet theorem, the topological genus is equal to the number of ``holes'' it has, for example, 0 for a sphere and 1 for a torus (a ``donut''). Just as the number of ``holes'' of a torus cannot be determined by local probing, $\mathcal{C}\mathpzc{h}$ is a global, ``topological'' property of a ground state manifold. To understand what determines topological transitions between its different integral values, we must consider that there are other energy levels above the ground state energy level $E_0$, and include the possibility of degeneracies where for example $E_0(\mathbf{R}) = E_1(\mathbf{R})$. In 4.2.1 we will see that degeneracies behave analogously to magnetic monopoles as the ``sources'' for $\mathbf{B}$ and, through the familiar Gauss's law, see in 4.2.2 that for well-behaved two-fold degeneracies $\mathcal{C}\mathpzc{h}$ simply counts the number of degeneracies enclosed by the manifold.

\subsubsection{\textcolor{SUBSUBSECTIONCOLOR}{4.2.1 Degeneracy as a source of Berry curvature}}
In this work, measurements are made of the ground state manifold $\ket{\psi_0(\mathbf{R})}$, but states of higher energy must be considered to understand the important role of degeneracy points in topological transitions. Namely, let $\ket{\psi_n(\mathbf{R})}$ denote the eigenstate corresponding to the $n^{\text{th}}$ energy level. To relate Berry curvature to degeneracy, we first use the fact that the curl of a gradient is zero along with the definitions of $\mathbf{A}$ and $\mathbf{B}$ [equations (\ref{eq:berry_connection}) and (\ref{eq:berry_curvature})] to write $\mathbf{B} = i\left[\nabla_\mathbf{R}\bra{\psi_0}\right]\times \left[\nabla_\mathbf{R}\ket{\psi_0}\right]$. We can then use the common trick of inserting the identity, expanded in terms of the energy eigenstates $\ket{\psi_n(\mathbf{R})}$, in between the bra and the ket: $\mathbf{B}(\mathbf{R})= i\sum_{n\ne 0} \left[\nabla_\mathbf{R} \bra{\psi_0}\right] \ket{\psi_n}\times \bra{\psi_n}\left[\nabla_\mathbf{R}\ket{\psi_0}\right]$, where we have excluded the $n=0$ term because it vanishes (this is easily seen as a consequence of normalization, $\braket{\psi_0}{\psi_0} = 1$). We can replace $\bra{\psi_n}\left[\nabla_\mathbf{R}\ket{\psi_0}\right]$ with the equivalent expression $\bra{\psi_n}\left[\nabla_\mathbf{R} \mathcal{H}\right]\ket{\psi_0}/(E_0 - E_n)$ for $n\ne 0$ (this is a straightforward consequence of differentiating the defining eigenvalue equation $\mathcal{H}(\mathbf{R}) \ket{\psi_n(\mathbf{R})} = E_n(\mathbf{R}) \ket{\psi_n(\mathbf{R})}$ with respect to $\mathbf{R}$ and rearranging terms), arriving at the equation
\begin{equation}
\mathbf{B}(\mathbf{R}) = i\sum_{n \ne 0} \frac{\bra{\psi_0}\left[\nabla_\mathbf{R} \mathcal{H}\right]\ket{\psi_n}\times \bra{\psi_n}\left[\nabla_\mathbf{R}\mathcal{H}\right] \ket{\psi_0}}{(E_n - E_0)^2}.\label{berry_curvature_2}
\end{equation}
From this, we can see that degeneracies (where $E_n = E_0$) can act as sources for the Berry curvature field $\mathbf{B}$. This also explicitly shows that $\mathbf{B}$ can be written without using phase-ambiguous derivatives $\nabla_\mathbf{R}\ket{\psi_0}$ of kets with respect to $\mathbf{R}$ [as in the definition of $\mathbf{A}(\mathbf{R})$], but instead in terms of more natural derivatives $\nabla_\mathbf{R} \mathcal{H}$ of $\mathcal{H}$ with respect to $\mathbf{R}$, meaning that it does not matter what phase we assign to eigenstates corresponding to different $\mathbf{R}$. Furthermore, we see that under certain assumptions about the behavior of $\mathcal{H}$ and $E_n$ near degeneracy, the singularities in Berry curvature are precisely the points of degeneracy. We also note that equation (\ref{berry_curvature_2}) relates the Berry curvature to the generalized force operator $-\nabla_\mathbf{R} \mathcal{H}$, which connects this discussion to formula (2) of the main text for the Lorentz force. A derivation of this force in terms of $\mathcal{B}_{\mu\nu}$ using perturbation theory can be found in \cite{GirtsevPNAS}.

\subsubsection{\textcolor{SUBSUBSECTIONCOLOR}{4.2.2 Sources of degeneracy as magnetic monopoles}}
Finally, we make the analogy between degeneracies and magnetic monopoles concrete. If we consider a closed 2D surface manifold $\mathcal{S}$ which bounds a 3D manifold in parameter space that possibly contains two-fold degeneracies, we can straightforwardly derive the interpretation of $\mathcal{C}\mathpzc{h}$ as the number of source ``magnetic monopole'' singularities enclosed by the ground state manifold. Note that we can assume in a 3D space that degeneracies will occur at isolated points \footnote{this is an instance of a general result \cite{vonneumann1929} stating that unless there is some special kind of symmetry in the Hamiltonian, three parameters must be tuned in order to reach a degeneracy}, and are therefore the magnetic monopoles that we seek. When only two energy levels $E_0$ and $E_1$ are involved in a two-fold degeneracy, we only need to consider one term from the sum (\ref{berry_curvature_2}) and can restrict ourselves to the relevant two-level subspace. It can be shown that $\mathcal{C}\mathpzc{h}$ is invariant under manifold perturbations as long as those perturbations don't cause a degeneracy to cross $\mathcal{S}$, so to extract the contribution to $\mathcal{C}\mathpzc{h}$ from a single enclosed degeneracy at $\mathbf{R}_0$ we are free to shrink the manifold down to a small sphere centered around $\mathbf{R}_0$ [so that only the $(E_1 - E_0)^2$ term contributes] and shift the origin of our coordinates to $\mathbf{R}_0$. With an appropriate rescaling of parameter space coordinates, following \cite{Berry1984} we can then write a general hermitian two-level Hamiltonian as
\begin{equation}
\mathcal{H} = \left( \begin{array}{cc} Z & X - iY \\ X + iY & -Z \end{array}\right),
\end{equation}
where $X$, $Y$, and $Z$ are the rescaled coordinates in the Pauli basis, i.e., $\mathbf{R} = (X,Y,Z)$ (the exact nature of this scaling is unimportant). In terms of this parametrization the energies are $E_{1/0}=\pm \sqrt{X^2 + Y^2 + Z^2} = \pm R$, so that the degeneracy is at the origin. We can immediately suspect that this leads to a monopole distribution for $\mathbf{B}$ because $1/(E_1 - E_0)^2\propto 1/R^2$. The precise calculation is dealt with in Berry's original paper \cite{Berry1984} using basic Pauli matrix algebra, resulting in the ground state Berry curvature field for a two-fold degeneracy at the origin,
\begin{equation}
\mathbf{B} = -\frac{\mathbf{R}}{2R^3}.
\end{equation}
We note that this is the same answer obtained for the Berry field for the specific case of a spin-$\frac{1}{2}$ particle subjected to a physical magnetic field \cite{Griffiths,Sakurai}. This is (up to a sign) the same expression for the magnetic field generated by a magnetic monopole of magnetic charge $1/2$, and therefore by Gauss's law leads to a contribution to $\mathcal{C}\mathpzc{h}$ of $(4\pi)/(2\pi)\times 1/2 = 1$, as we claimed. Gauss's law then immediately yields for our experiment
\begin{equation}
\mathcal{C}\mathpzc{h} = Q^\text{enc}_\text{m}.
\end{equation}

\noindent
\subsection{\textcolor{SUBSECTIONCOLOR}{4.3 Choice of coordinate system}}

Here we clarify the choice of coordinate system used throughout the main work. There is some ambiguity in how we define the Berry connection in spherical coordinates.  One way is to close our eyes and pretend that we don't know that $\theta$ and $\phi$ are spherical angles, instead simply treating them as Euclidean parameters. We will call this the ``Cartesian'' choice, which gives for example the $\phi$-component
\begin{equation}
A_\phi^C = i \braxket{\psi_0}{\partial_\phi}{\psi_0}.
\end{equation}
Alternatively, we could explicitly take into account the non-Euclidean metric associated with spherical coordinates, using $\nabla f = \frac{\partial f}{\partial r} \hat r + \frac{1}{r} \frac{\partial f}{\partial \theta} \hat \theta + \frac{1}{r \sin \theta} \frac{\partial f}{\partial \phi} \hat \phi$ \cite{Jackson} to yield the ``spherical'' definition
\begin{equation}
A_\phi^S = i \frac{1}{r \sin \theta} \braxket{\psi_0}{\partial_\phi}{\psi_0} = \frac{A_\phi^C}{r \sin \theta} ~.
\end{equation}
This difference may have confused the reader. Below, we will show that either method works, and both of them can be used to arrive at Eq. (4) of the main text.

\subsubsection{\textcolor{SUBSUBSECTIONCOLOR}{4.3.1 Cartesian coordinates}}
This is arguably the simpler method, though it is a bit harder to justify.  As above, for the ground state manifold we define for $\phi$ or $\theta$
\begin{equation}
A_{\phi/\theta}^C \equiv i \braxket{\psi_0}{\partial_{\phi/\theta}}{\psi_0}.
\end{equation}
Since we are integrating over a spherical surface ($r$ fixed), we will not need to take any derivatives with respect to $r$.  In these pseudo-Cartesian coordinates then, the non-trivial component of the Berry curvature is
\begin{equation}
B_{\theta \phi}^C \equiv \partial_\theta A_\phi^C - \partial_\phi A_\theta^C ~.
\end{equation}
We can then perform the surface integral by noting that, in Cartesian coordinates, the surface element is just $dS = d\theta d\phi$, so that for a spherical manifold
\begin{equation}
\mathcal{C}\mathpzc{h}  = \frac{1}{2\pi} \int {\bf B} \cdot d{\bf S} = \frac{1}{2\pi} \int_0^{2\pi} d\phi \int_0^\pi d\theta B_{\theta \phi}^C ~,
\label{eq:Ch_c}
\end{equation}
which is the expression we expected.

\subsubsection{\textcolor{SUBSUBSECTIONCOLOR}{4.3.2 Spherical coordinates}} If we take the spherical version of the gradient, then the $\phi$ and $\theta$ components are
\begin{align}
A_\phi^S &= i \frac{1}{r \sin \theta} \braxket{\psi_0}{\partial_\phi}{\psi_0} = \frac{A_\phi^C}{r \sin \theta}~, \notag \\
A_\theta^S &= i \frac{1}{r} \braxket{\psi_0}{\partial_\theta}{\psi_0}  = \frac{A_\theta^C}{r}~.
\end{align}
The Berry curvature vector is given by ${\bf B}^S = \nabla \times {\bf A}^S$, which in general is a complicated expression.  However, for our spherical surface of integration, the Chern integral is given by
\begin{equation}
\mathcal{C}\mathpzc{h} = \frac{1}{2\pi} \int {\bf B} \cdot d{\bf S} = \frac{1}{2\pi} \int B_r^S d S_r ~,
\label{eq:Ch_s}
\end{equation}
since the surface element is strictly radial:
\begin{equation}
d{\bf S} = \hat r d S_r = \hat r (r^2 \sin \theta d \theta d \phi) ~,
\label{eq:dS}
\end{equation}
where we have used the standard form of a spherical surface element.  Taking the curl in spherical coordinates, the radial component of ${\bf B}^S$  is
\begin{equation}
B_r^S = \frac{1}{r \sin \theta} \left[ \partial_\theta(\sin \theta A_\phi^S) - \partial_{\phi} A_\theta^S \right] = \frac{1}{r^2 \sin \theta} \left[ \partial_\theta A_\phi^C - \partial_{\phi} A_\theta^C \right] = \frac{B_{\theta \phi}^C}{r^2 \sin \theta}~.
\label{eq:Fr}
\end{equation}
Plugging Eqs. (\ref{eq:dS}) and (\ref{eq:Fr}) into (\ref{eq:Ch_s}), we again get the Cartesian expression for $\mathcal{C}\mathpzc{h}$ (\ref{eq:Ch_c}).

Finally, we note that for our case, the Hamiltonian is cylindrically invariant: we can get the Hamiltonian at arbitrary $\phi$ from the Hamiltonian at $\phi=0$ by just rotating the spins by an angle $\phi$ around the $z$-axis. Accordingly, the Berry curvature must be cylindrically symmetric, meaning that $B_{\theta\phi}(\theta,\phi)=B_{\theta\phi}(\theta)$ is independent of $\phi$.  Therefore, if we plug into the expression for the Chern number, we find
\begin{equation}
\mathcal{C}\mathpzc{h} = \underbrace{\frac{1}{2\pi} \int_0^{2\pi} d\phi}_{=1} \int_0^\pi d\theta B_{\theta \phi}(\theta) = \int_0^\pi B_{\theta \phi}(\theta) d\theta~.
\end{equation}

We now show how the equation (4) of the main text was derived. Starting with the Hamiltonian of a single qubit or equivalently spin-1/2 particle in a magnetic field:

\begin{equation}\label{singlemag}
\mathcal{H}_S = -\frac{\hbar}{2}(H_X\sigma^x + H_Y\sigma^y + H_Z \sigma^z),
\end{equation}
and re-parameterizing it for spherical coordinates, it becomes

\begin{equation}
\mathcal{H}_{S}(H_{r},\theta,\phi)=-\frac{\hbar}{2}H_{r}(\sin\theta\cos\phi\,\sigma^{x}+\sin\theta\sin\phi\,\sigma^{y}+\cos\theta\,\sigma^{z}).\label{sm}
\end{equation}

Therefore,
\begin{equation}
F_{\phi}=-\left\langle\partial_\phi \mathcal{H}(\phi=0)\right\rangle
=\frac{\hbar}{2}H_{r}\sin\theta\,\langle\sigma^{y}\rangle.\label{sm}
\end{equation}

Using equation (2) of the main text,
\begin{equation}
\hbar B_{\theta \phi} d\theta=\frac{\hbar}{2}H_{r}\sin\theta\,\langle\sigma^{y}\rangle dt,
\end{equation}

which is used in the main text in computing $\mathcal{C}\mathpzc{h}$ from the measured values of \(H_r\) and \(\langle\sigma^{y}\rangle\).

\section{\textcolor{SECTIONCOLOR}{5. Mapping the Two-Qubit Hamiltonian to Electronic Band Structure} }

As in the main text, we consider the two-qubit Hamiltonian
\begin{equation}
\mathcal{H}=-\frac{\hbar}{2} \Big[ H_0 \sigma^z_1 + H_r \hat n(\theta,\phi) \cdot ({\bf \sigma_1} + {\bf \sigma_2}) - g (\sigma^x_1 \sigma^x_2 + \sigma^y_1 \sigma^y_2 ) \Big]
\label{eq:H_2q}
\end{equation}
for fixed $H_0$, $H_r$, and $g$. For this section we assume \(\hbar=1\). At a given value of $\theta$ and $\phi$,  this Hamiltonian is a $4 \times 4$ matrix; a general $N$-qubit Hamiltonian would similarly be $2^N \times 2^N$.  To help understand the topology of this Hamiltonian, we wish to map it to a more conventional electronic Hamiltonian, as we did in mapping the single qubit to the Haldane model of graphene.  In this supplement, we show that \eqref{eq:H_2q} can be mapped to either a four-band model of non-interacting electrons in the spirit of the Haldane mapping or a four-band interacting electron model with interactions that are short-range in momentum space.  Finally, we comment on the extension of these mappings to higher numbers of qubits.

For both non-interacting and interacting electron mapping, we again utilize the idea that a given angle $(\theta,\phi)$ of the rotating field $H_r$ corresponds to a point in momentum space (see the single qubit Haldane supplement): ${\bf k} = (k_x,k_y) \leftrightarrow (\theta,\phi)$.  Then the simple idea which worked for mapping the single qubit to the Haldane model is to ``fermionize'' the spin:
\begin{equation}
\sigma^\alpha_j \to \sum_{s s'} c^\dagger_{js} \sigma^\alpha_{s s'} c_{j s'} ~,
\end{equation}
where $\alpha=x,y,z$, $j=1,2$ specifies the qubit, and $s,s' = \{\uparrow, \downarrow\}$ iterate through the spin states.  For example, this mapping gives $\sigma^x_1 \to c_{1\uparrow}^\dagger c_{1\downarrow} + c_{1\downarrow}^\dagger c_{1\uparrow}$.  Performing these replacements we get

\begin{eqnarray}
\nn \mathcal{H}({\bf k}) & = & (H_r \cos \theta_{\bf k} + H_0)[\underbrace{(c^{\bf k}_{1 \uparrow})^\dagger c^{\bf k}_{1 \uparrow} - (c^{\bf k}_{1 \downarrow})^\dagger c^{\bf k}_{1 \downarrow}}_{\sigma^z_1}] + H_r \cos \theta_{\bf k} [\underbrace{(c^{\bf k}_{2 \uparrow})^\dagger c^{\bf k}_{2 \uparrow} - (c^{\bf k}_{2 \downarrow})^\dagger c^{\bf k}_{2 \downarrow}}_{\sigma^z_2}] +
\\\nn&&H_r \sin \theta_{\bf k} \cos \phi_{\bf k} [\underbrace{(c^{\bf k}_{1 \uparrow})^\dagger c^{\bf k}_{1 \downarrow} + (c^{\bf k}_{1 \downarrow})^\dagger c^{\bf k}_{1 \uparrow}}_{\sigma^x_1} + (1 \to 2)] +
\\\nn&& H_r \sin \theta_{\bf k} \sin \phi_{\bf k} [\underbrace{-i (c^{\bf k}_{1 \uparrow})^\dagger c^{\bf k}_{1 \downarrow} +i (c^{\bf k}_{1 \downarrow})^\dagger c^{\bf k}_{1 \uparrow}}_{\sigma^y_1} + (1 \to 2)] +
\\&&\frac{g}{2} [(c^{\bf k}_{1 \uparrow})^\dagger c^{\bf k}_{1 \downarrow} (c^{\bf k}_{2 \downarrow})^\dagger c^{\bf k}_{2 \uparrow} + (\uparrow \leftrightarrow \downarrow)] ~.
\end{eqnarray}
The last term of this Hamiltonian contains a four-fermion operator, so this is an interacting fermionic Hamiltonian with four flavors of fermion ($c_{1\uparrow}, c_{1\downarrow},c_{2\uparrow}, c_{2\downarrow}$).  To maintain one spin per qubit, we want the many-body ground state at half-filling and without double occupancy on ``site'' $j=1,2$.  However, the interaction remains short-range in momentum space, meaning the electronic Hamiltonian is still separable into momentum sectors: $H=\sum_{\bf k} H_{\bf k}$.  Such models are similar to the mean-field BCS Hamiltonian \cite{BCS1957}, in this case with the additional wrinkle of being local in momentum space.
momentum space.

\begin{figure}
\includegraphics[width=.75\linewidth]{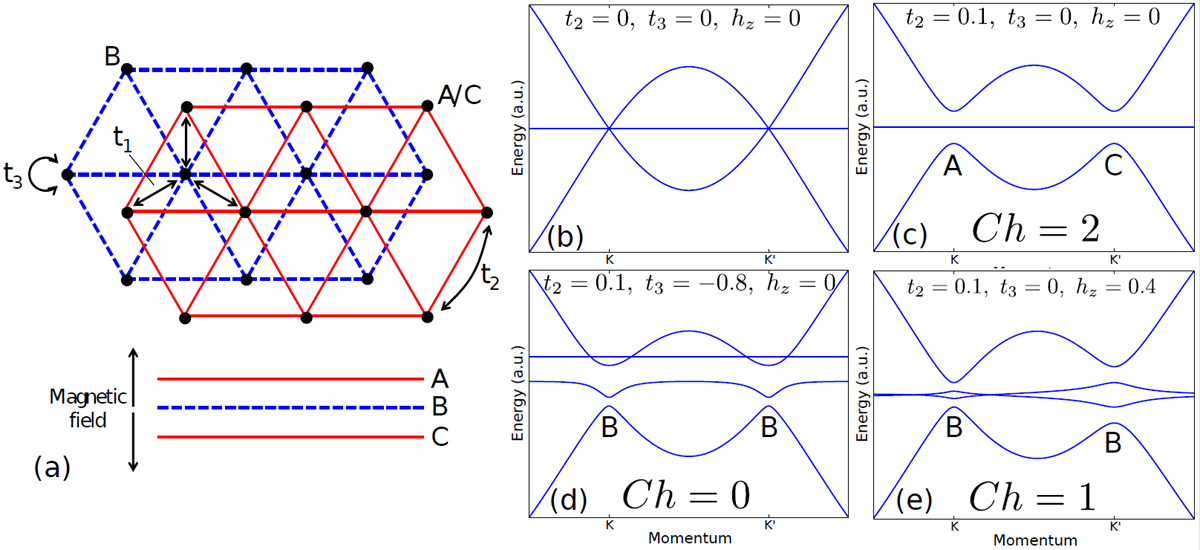}
\caption{Illustration of the four-band non-interacting lattice model to which we map our two-qubit model. \textbf{(a)} The model consists of three stacked triangular lattices (A, B, and C), the middle of which (B) contains two spin/orbital states.  In addition to nearest neighbor hopping ($t_1$) and on-site hybridization the B sublattice ($t_2$), electrons on the A and C sublattices experience magnetic field that adds phase to the hopping ($t_2 e^{i \varphi}$).  Finally, an effective Zeeman field splits the spin/orbital states on all sublattices.  \textbf{(b)} to \textbf{(e)} Energy dispersions for this model along a cut containing the $K$ and $K'$ corners of the first Brillouin zone.  We fix energy by setting $t_1=1$.  At the $K$ and $K'$ points, the sublattices decouple; we label the sublattice that is occupied in the ground state at these points.  $\mathcal{C}\mathpzc{h}$ then counts the number of times the wavefunction ``twists'' between the sublattices.}
\label{fig:haldane_2qubit}
\end{figure}

While this first mapping is true to the interacting nature of the qubit, it gives little physical insight into the topological transition.  To try to understand this better, we now discuss how the same system can be mapped to a four-band Haldane-like model of non-interacting electrons.  Unlike the interacting case, we present a microsopic model that will realize this topology.  The model is shown schematically in \fref{fig:haldane_2qubit}a
. The basic idea is to consider electrons hopping on stacked triangular lattices with a single internal degree of freedom (spin/orbital/etc.) that can take one of two values, which we denote $\uparrow$ and $\downarrow$.  The middle layer of the stack, which we call B, supports both $\uparrow$ and $\downarrow$ states, while the upper (lower) layer supports only $\uparrow$ ($\downarrow$).  This could be realized, for example, by a lattice where the middle layer has two orbital states (e.g. $p_x$ and $p_y$ orbitals), while the outer layers have only one orbital state (e.g. $s$ orbitals).  We assume there is a magnetic field gradient as in \fref{fig:haldane_2qubit}a,
which gives zero field layer B and yields opposite magnetic field at A and C.  We then consider four quadratic terms in the Hamiltonian:

\begin{enumerate}
\item Nearest neighbor hopping $t_1$, which connects the B sublattice to the A and C sublattices.  The matrix element is assumed to be equal for spin up hopping to either up or down, which make senses for orbital degrees of freedom or if $\uparrow$ and $\downarrow$ represent real spins, but with different quantization axes on A/C than on B.
\item Second neighbor hopping on the A and C sublattices, which picks up a phase due to the magnetic flux.  For simplicity, we consider a flux of $\Phi=3 \Phi_0/4 = -\Phi_0/4$ per plaquette, where $\Phi_0=h/e$ is the quantum of flux; this gives phase $\varphi=\pi/2$ on the hopping, removing a diagonal shift in the energy bands of the A and C sublattices (see \cite{Haldane1988}).
\item On-site hybridization $t_3$ between the effective spin states on the B sublattice.
\item An effective Zeeman shift $h_z s^z$, where $s^z$ is the internal spin/orbital degree of freedom.
\end{enumerate}

Let us examine these terms in the language of the Haldane Hamiltonian.  First note that the A and B (or equivalently C and B) sublattices have the exact
structure as the sublattices of monolayer graphene.  Therefore, the first Brillouin zone of this non-interacting electron model is equivalent to that of
graphene, and we can naturally expand the Hamiltonian in small deviations of the momenta around the non-identical zone corners $K$ and $K'$ (see [cite
Haldane supplement]).  There are four states in each unit cell: $B\uparrow$, $B\downarrow$, $A\uparrow$, and $C\downarrow$.  The nearest neighbor hopping
$t_1$ is the only term that connects the sublattices, so it is responsible for producing a graphene-like dispersion relation.  However, to get the Chern
number of 2, this is slightly different from the Haldane model of graphene.  To see this, consider quantizing the spins along the $x$-axis.  It is easy to see
that the state \(\ket{\uparrow_x} = \frac{1}{\sqrt 2} (\ket{\uparrow_z} +\ket{\downarrow_z}\)
hops freely and will give the dispersion of a graphene lattice with hopping amplitude $2 t_1$.  However, a state with spin \(\ket{\downarrow_x} = \frac{1}{\sqrt 2} (\ket{\uparrow_z}- \ket{\downarrow_z}\) is annihilated by this hopping term, so in addition to the graphene dispersion, there are two flat bands at energy zero if only the $t_1$ term is considered
(see \fref{fig:haldane_2qubit}(b)).

The remaining terms then determine the topology by breaking the degeneracies at momenta $K$ and $K'$.  For instance, the $t_2 e^{i\varphi}$ hopping only occurs on the A and C sublattices, so at the $K$ and $K'$ points (where the $t_1$ hopping vanishes), the electrons only live on the A or the C sublattice.  As the momentum is varied from $K$ ($\theta=0$) to $K'$ ($\theta=\pi$), the electronic ground state winds from sublattice A to C, which results in $\mathcal{C}\mathpzc{h}=2$ (see \fref{fig:haldane_2qubit}(c)).
This is precisely the action of the probe field, so we see that $t_2 \sim H_r$.  Similarly, the $t_3$ term hybridizes the orbitals on the B lattice, causing the energy of the symmetric state on the B lattice to go down.  For strong enough $t_3$, this can push the energy of the symmetric state on B lattice below the A and C energies throughout the Brillouin zone, resulting in Chern number zero (i.e., no wrapping of wave function, see \fref{fig:haldane_2qubit}(d)).
  This is the same role as the qubit interactions, so not surprisingly $t_3 \sim g$.  Finally, if we again consider $t_3=0$, then a large positive Zeeman field $h_z$ will push the energy of the spin down state below that of the spin up.  In the presence of $t_2$ hopping, this gives a ground state winding from the A sublattice to the B sublattice as momentum goes from $K$ to $K'$ (\fref{fig:haldane_2qubit}(e)), yielding Chern number one.  Not surprisingly, this gives that $h_z \sim H_0$.

More explicitly, the Hamiltonian described above can be written
\begin{eqnarray}
\nn \mathcal{H} &=& \sum_{\bf r} \big[ -t_1 \sum_j (c_{{\bf r}\up}^\dagger + c_{{\bf r}\up}^\dagger)(c_{{\bf r + a_j}\down} + c_{{\bf r+a_j}\down}^\dagger) - t_2  \sum_j ( \overbrace{e^{i\varphi} c_{{\bf r}\up}^\dagger c_{{\bf r+b_j}\up}}^\mathrm{A~sublattice} + \overbrace{e^{-i\varphi} c_{{\bf r}\down}^\dagger c_{{\bf r+b_j}\down}}^\mathrm{C~sublattice}) +
\\ && \frac{t_3}{2} (c_{{\bf r+a_1}\up}^\dagger c_{{\bf r+a_1}\down} + (\up \leftrightarrow \down)) - \frac{h_z}{2} (c_{{\bf r} \up}^\dagger c_{{\bf r} \up} + c_{{\bf r+a_1} \up}^\dagger c_{{\bf r+a_1} \up}  - (\up \to \down) ) \big] + h.c. ~,
\end{eqnarray}
where ${\bf r}$ are the sites on the A/C sublattice, ${\bf a_j}$ are the nearest neighbor displacements, and (following Haldane's convention), ${\bf b_j}$ are the next-nearest-neighbor displacements along directions with positive hopping phase on the A sublattice.  Diagonalizing with phase $\varphi=\pi/2$, this gives Bloch Hamiltonian
\begin{equation} \label{david8}
\mathcal{H}_{\bf k}=
\begin{pmatrix}
-2 t_2 \sum_j \sin({\bf k \cdot b_j}) - h_z &
-2 t_1 \sum_j \cos({\bf k \cdot a_j}) &
-2 t_1 \sum_j \cos({\bf k \cdot a_j}) &
0 \\
-2 t_1 \sum_j \cos({\bf k \cdot a_j}) &
-h_z &
t_3 &
-2 t_1 \sum_j \cos({\bf k \cdot a_j}) \\
-2 t_1 \sum_j \cos({\bf k \cdot a_j}) &
t_3 &
h_z &
-2 t_1 \sum_j \cos({\bf k \cdot a_j}) \\
0 &
-2 t_1 \sum_j \cos({\bf k \cdot a_j}) &
-2 t_1 \sum_j \cos({\bf k \cdot a_j}) &
2 t_2 \sum_j \sin({\bf k \cdot b_j}) + h_z
\end{pmatrix},
\end{equation}
where the columns denote A$\up$, B$\up$, B$\down$, and C$\down$ in that order.  For comparison, the two-qubit Hamiltonian in the basis $\up \up$, $\up \down$, $\down \up$, $\down \down$
\begin{equation}
\mathcal{H}_{2Q} = \frac{1}{2} \left(\begin{array}{cccc}
-2 H_r \cos \theta - H_0 & -2 H_r \sin \theta &  -2 H_r \sin \theta & 0
\\ -2 H_r \sin \theta & -H_0 & -2g &  -2 H_r \sin \theta
\\ -2 H_r \sin \theta & -2g & H_0 &  -2 H_r \sin \theta
\\ 0 &  -2 H_r \sin \theta &  -2 H_r \sin \theta & 2 H_r + H_0
\end{array}\right) ~.
\end{equation}

By inspecting these two Hamiltonians, we see that they map to each other under the identification
\begin{equation}
{\bf k} \leftrightarrow (\theta,\phi) ~,~ 3 t_2 \sqrt 3 = H_r ~,~ -t_3 = g ~,~ 2 h_z = H_0 ~,
\end{equation}
where we used the fact that $\sum_j \sin({\bf k}\cdot{\bf b_j}) = \pm 3 \sqrt 3 / 2$ at the corners of the first Brillouin zone. Therefore, the topology of the ground band of this four-band electronic model is equivalent to that of the two-qubit system that we experimentally investigate.

It is clear from the above discussion that a system of $L$ qubits with $2^L$ eigen states would map to a non-interacting model with $2^L$ bands.  While 2 or 4 band models are not so crazy, an eight band model with only a singled filled band -- as would be needed for $L=3$ qubits -- is starting to get physically less realistic.  Clearly the scaling of the number of bands with the number of qubits is such that these non-interacting Haldane-like models will become exponentially more difficult to engineer as the system becomes larger. Working instead with the interacting model helps quite a bit; simple counting requires only $2L$ flavors of fermion (spin up and down for each qubit) at half-filling and with no double-occupancy.   However, this model has no obvious microscopic interpretation, so for the time being we consider it less physical.  Therefore, we conclude that as the qubit number is increased (and restricted to the above mapping methods), it becomes increasingly unworkable to think of the system in terms of electrons on a lattice.  For large spin lattices, we really should think of our measurement as simply probing the topology of the spin manifold, a problem which is interesting in and of itself. It is also worth pointing out that the two mapping we have described about only work for our choice of parameter manifolds, namely fixed external field strength with a rotating angle applied equally to each qubit. By using different choices of manifold, even within the same two-qubit system, we can engineer different effective condensed matter models, demonstrating the flexibility of these two-qubit systems.

\clearpage
\section{\textcolor{SECTIONCOLOR}{6. Experimental protocols, calibration, and analysis}}

Here we provide the outline of the experiment and its basic protocols. The first step is the calibration of the pulses so we know \(H_{X}/2\pi\) and \(H_{Z}/2\pi\) with good accuracy. An important aspect of calibration is also finding the compensating pulse such that when we only \(H_{X}/2\pi\), the state of the qubit remains in the YZ plane. The details of these steps are explained in Fig. S5. Next, one needs to find a proper ramp speeds to be sure the higher order errors in equation 2 of main text remain small; another words, how much non-adiabatic a ramp can be and still yield a good result. This is shown in figure 6, where we explored the three parameters that set the non-adiabaticity of a ramp: \(H_{X}/2\pi\), \(H_{Z}/2\pi\), and \(T_f\). After finding that one needs to set all the ramps such that the adiabaticity measure $A$ remains acceptable. After finding proper ramp speeds and calibrations, one can do the single qubit experiment, which involves applying \(H_{X}/2\pi\) with sine envelope and \(H_{Z}/2\pi\) with a cosine profile, as discussed in Fig. S4. The two qubit experiment requires additional calibrations to what is mentioned in here and is discussed in \cite{Yu2014}.

\begin{figure*}
\begin{centering}
\includegraphics[width=150mm]{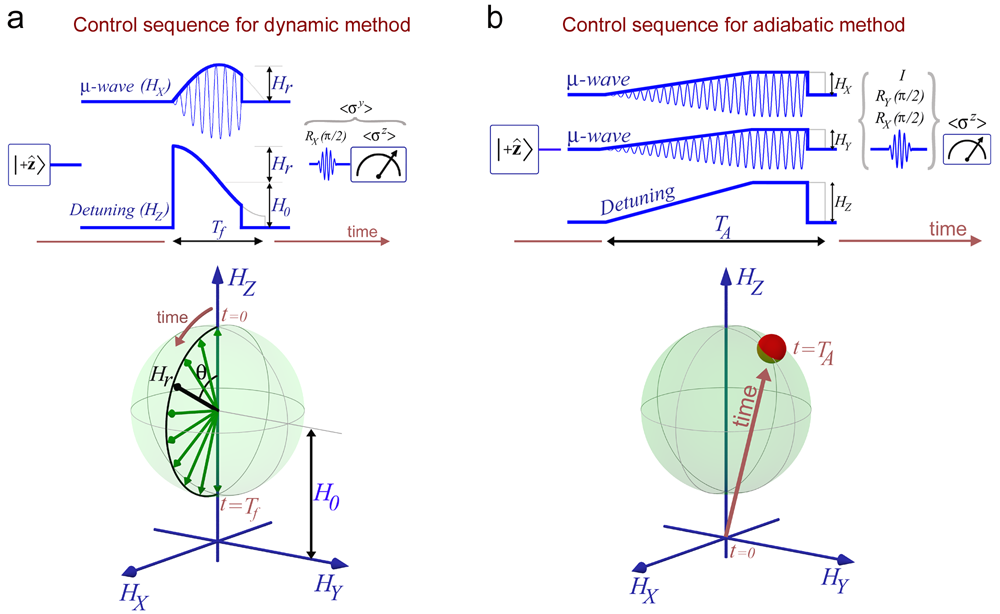}
\par\end{centering}
\caption{
\textbf{Control sequence used for the single qubit experiments.} \textbf{(a)} The pulse sequence used to obtain the phase diagram shown in Fig. 3(a) of the main text. Every control sequence began with preparing the qubit in its ground state, which was achieved by waiting for times much longer than the qubit relaxation time (a few tens of microseconds). In the phase diagram measurements, \(\left<\sigma^y\right>\) was measured at $50$ time steps during the ramps, where the first data point was measured at $t=0$ and the last one at $t=T_f=1000$ ns. To measure each data point the sequence was repeated $300$ times. This $50$-point \(\left<\sigma^y\right>\) profile as a function of time from $0$ to $T_f$ was then multiplied by a sine profile (see equation (4) in the main text) and integrated to give $\mathcal{C}\mathpzc{h}$. \(\left<\sigma^y\right>\) was measured by inserting a $R_X(\pi/2)$ pulse before the \(\left<\sigma^z\right>\) measurement. The microwave pulse with a sine profile and detuning pulse with a cosine profile constitute a semi-circular ramp in the parameter space, and given the symmetry of the single-qubit Hamiltonian, this is sufficient to calculate the curvature over the entire spherical manifold. \textbf{(b)} Control sequence for adiabatic state preparation and measurement. In contrast to the dynamic method (equation (4) of the main text), in the adiabatic state preparation process, the qubit needs to remain close to the instantaneous eigenstate of the system during the ramp. To evolve to the ground state of the Hamiltonian with parameters \((H_X,H_Y,H_Z)\), we start from the origin of parameter space, where all pulses are zero, and gradually turn \((H_X,H_Y,H_Z)\) to their final values in 500 ns. The pulses then remain at their target values for 500\,ns (hence \(T_A=1000\,ns\)). Over this fixed pulse regions at 100 points (distributed uniformly from 500\,ns to 1000\,ns) the state of the qubit was measured with tomography and the results are averaged to present a single Bloch vector data corresponding to given \((H_X,H_Y,H_Z)\) values. To visualize the ground states over the entire spherical manifold $\mathcal{S}$, the process was repeated for different values of \((H_X,H_Y,H_Z)\) to form a grid over this parameter space  sphere.}
\label{fig:concept}
\end{figure*}

\begin{figure*}
\begin{centering}
\includegraphics[width=140mm]{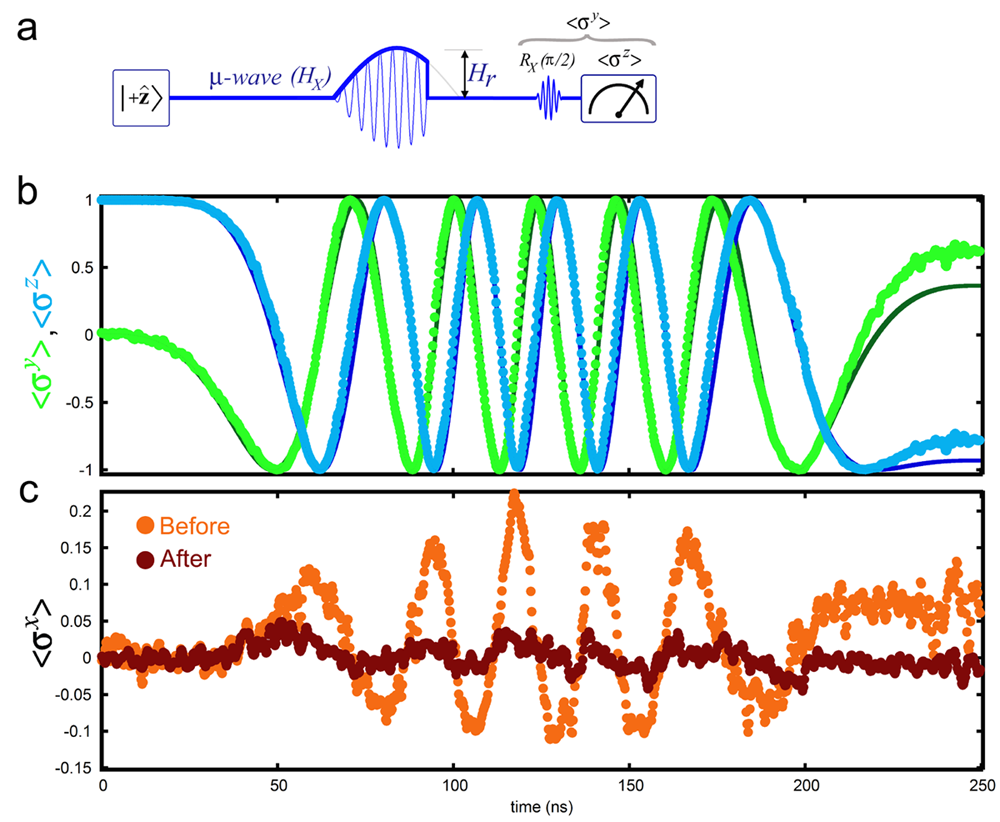}
\par\end{centering}
\caption{
\textbf{Pulse Calibration.} Single qubit microwave and detuning pulses were calibrated separately before applying them simultaneously to the qubits. \textbf{(a)} To calibrate the microwave pulse, the qubit was prepared in its groundstate \(|+\hat{z}\rangle\) and a microwave pulse with a sine envelope of amplitude \(H_r\) was applied to the qubit. The state of the qubit (the Bloch vector) was measured at each point in time by interrupting the ramp and performing full state tomography. As shown in panel \textbf{(a)}, for instance, to measure \(\left<\sigma^y\right>\), a rotation of \(\pi/2\) around the X axis was performed before the \(\left<\sigma^z\right>\) measurement. This pulse results in a cyclic motion of the Bloch vector in the Y-Z plane, with a non-zero out of plane component. The out of plane component is mainly due to leakage to other states due to finite inharmonicity of the qubit system. Therefore, the measured out of plane component (\(\left<\sigma^x\right>\), orange points in panel \textbf{(c)} ) needs to be calibrated, which was done by adding a compensating microwave pulse on the Y-axis, with a variable amplitude during the pulse sequence such that it keeps  \(\left<\sigma^x\right>\) close to zero. A typical result before and after calibration is shown in \textbf{(c)}. Fitting the \(\left<\sigma^z\right>\) and \(\left<\sigma^y\right>\) with a single fitting parameter can be done using the Schr\"{o}dinger equation. The resulting value in this case is \(H_{r}/2\pi=44.5\) MHz. The dark blue and green solid lines are the result of the fitting. During the calibration since full state tomography was performed, we normalized the measured values of \(\left<\sigma^x\right>\), \(\left<\sigma^y\right>\), \(\left<\sigma^z\right>\) such that \(\left<\sigma^x\right>^2+\left<\sigma^y\right>^2+\left<\sigma^z\right>^2=1\). The detuning pulse was applied and measured similarly. This was done by bringing the qubit to the equator of the Bloch sphere with a \(\pi/2\)-pulse first, and then applying the detuning in the absence of microwave pulse, and fitting the result with the Schr\"{o}dinger equation. There was no compensation pulse to be considered in this case. }
\label{fig:concept}
\end{figure*}

\begin{figure*}
\begin{centering}
\includegraphics[width=170mm]{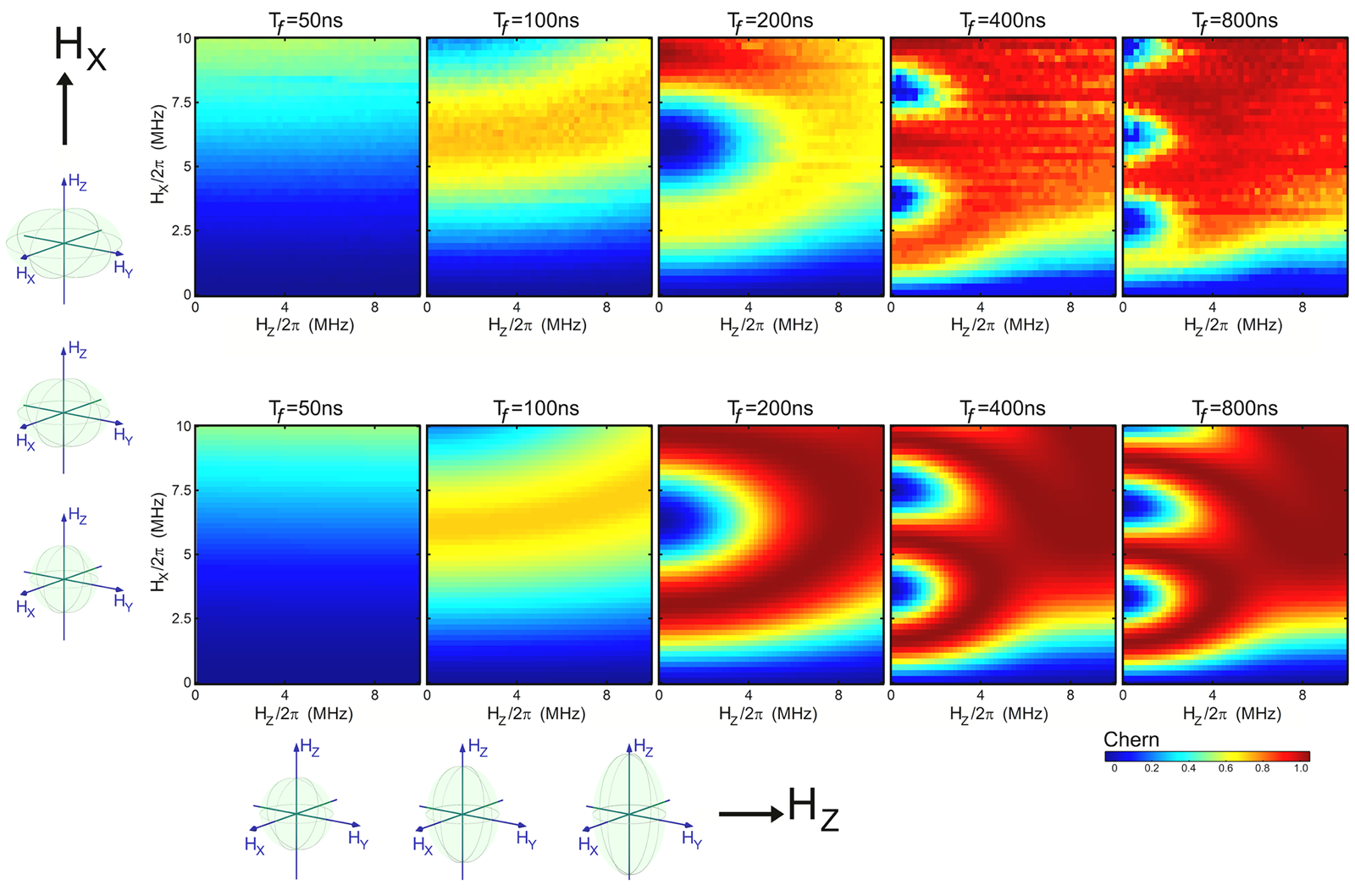}
\par\end{centering}
\caption{
\textbf{Adiabaticity required to measure $\mathcal{C}\mathpzc{h}$.} Although equation (2) of the main text does not require adiabaticity, it does require the ramp in parameter space to be done slowly, such that \(O(v_\theta^2)\) remains negligible. One needs to operate with ramp velocities for which deviation from adiabaticity varies linearly with the ramp speed. Slower ramps are more adiabatic and hence better in this regard, but they have a small deviation from adiabaticity, which would be hard to measure experimentally. On the other hand, ramps that are too fast also contain non-adiabatic errors that are not linearly proportional to ramp speed, and hence should be avoided. In this figure, the top row shows the experimental results of measuring $\mathcal{C}\mathpzc{h}$ by making various elliptical ramps and traversing them with different velocities. A microwave pulse of \(X(t)= H_{X}\sin(\pi t/T_f)\) and detuning of \(Z(t)= H_{Z}\cos(\pi t/T_f)\) are used, with \(H_{X}/2\pi\) and \(H_{Z}/2\pi\) varied from 0 to 10 MHz. Five different speeds are used, which are set by \(T_f\), where \(T_f\) is the time it takes to ramp from the north pole of the manifold to its south pole. The lower row shows the numerical results using the same ramps, obtained from the time dependent Schr\"{o}dinger equation. In this example, we seek to measure $\mathcal{C}\mathpzc{h}$ over a manifold of ground states that encloses the origin of the parameter space. The theoretical value of $\mathcal{C}\mathpzc{h}$ in this case is 1 \cite{Griffiths}. From left to right, as \(T_f\) becomes longer, the ramps are more adiabatic and the measured value for $\mathcal{C}\mathpzc{h}$ approaches one. In each panel, moving from lower left to upper right, adiabaticity increases, since \(A=T_fH_r/2\pi=T_f\sqrt{{H_X}^2+{H_Z}^2}/2\pi\). For $T_f=400$ ns or longer, a good estimate of $\mathcal{C}\mathpzc{h}$ can be achieved, as almost the entire plot is red, regardless of the shape of the manifold. The method yields a good estimate of $\mathcal{C}\mathpzc{h}$ for \(A>1.5\). To provide a visual guide, the deformation of the spherical manifolds to ellipsoids, by keeping \(H_X\) fixed and increasing \(H_Z\) (horizontal axis below figure), and by keeping \(H_Z\) fixed and increasing \(H_X\) (vertical axis left of figure) are shown. }
\label{fig:concept}
\end{figure*}

\begin{figure*}
\begin{centering}
\includegraphics[width=100mm]{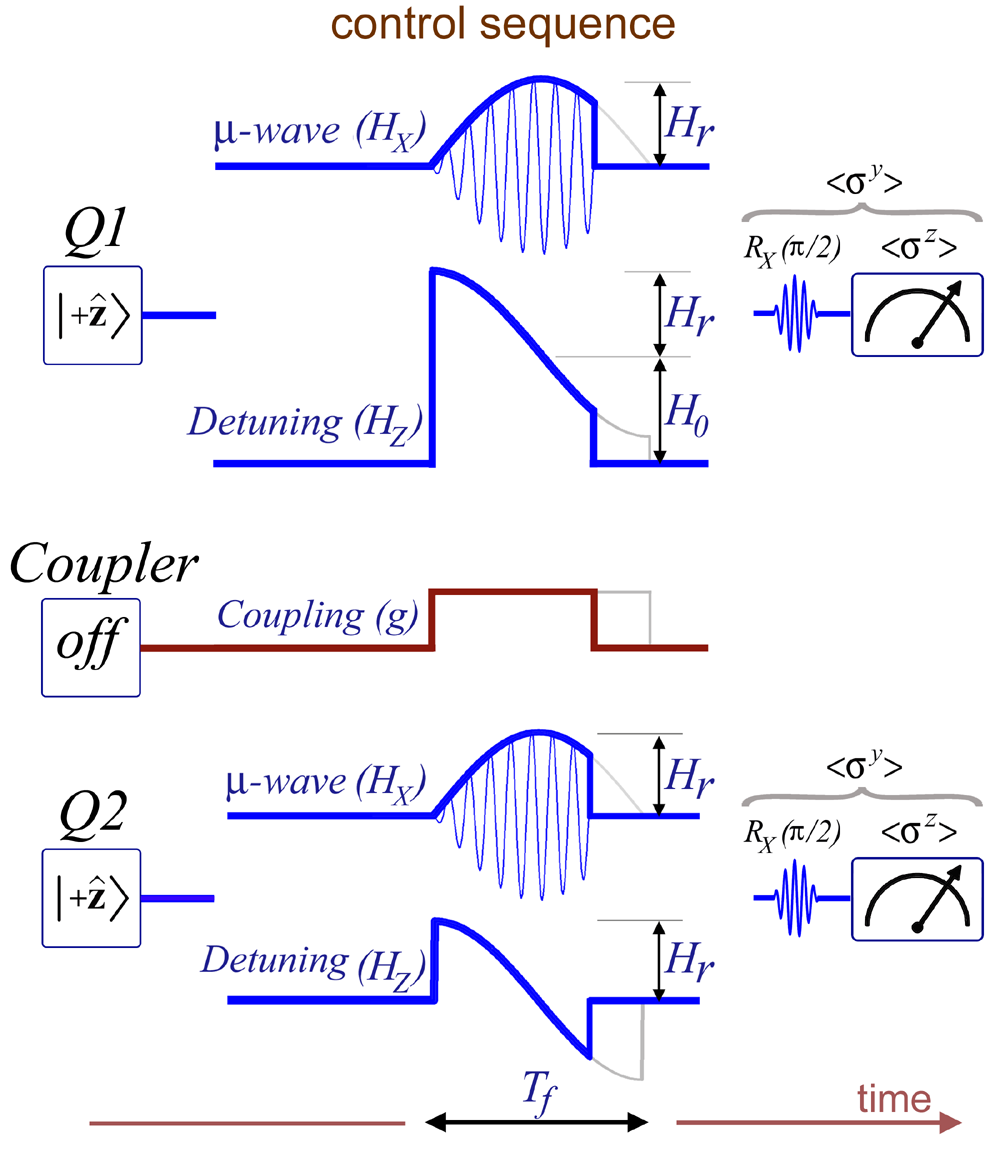}
\par\end{centering}
\caption{
\textbf{Control sequence used for the 2-qubit experiments.} \textbf{(a)} The pulse sequence used on individual qubits and the coupler elements are shown. The microwave pulse applied to each qubit had a sine envelope. The detuning of Q2 has a cosine profile, and the detuning of Q1 has a cosine profile plus an offset defining $H_0$. With a rectangular pulse, the coupling between the two qubits is turned on during the active part of other pulses. The synchronization of the pulses as well as finding the flux value corresponding to \(g=0\) were done \cite{Yu2014} prior to running the sequence. In addition, a calibration matrix to take various types of crosstalk into account was measured and implemented. This included both microwave and flux-biasing crosstalk \cite{Yu2014}. Using equation (4) of the main text, the $\mathcal{C}\mathpzc{h}$ for 2-qubit manifolds is the summation of individual ones. Therefore, each pulse sequence was run twice, once to measure \(\left<\sigma^y_1\right>\) and again to measure \(\left<\sigma^y_2\right>\) and the results were added to give the phase diagram plots shown in the Fig. 4 of the main text.}
\label{fig:concept}
\end{figure*}

\begin{figure*}
\begin{centering}
\includegraphics[width=170mm]{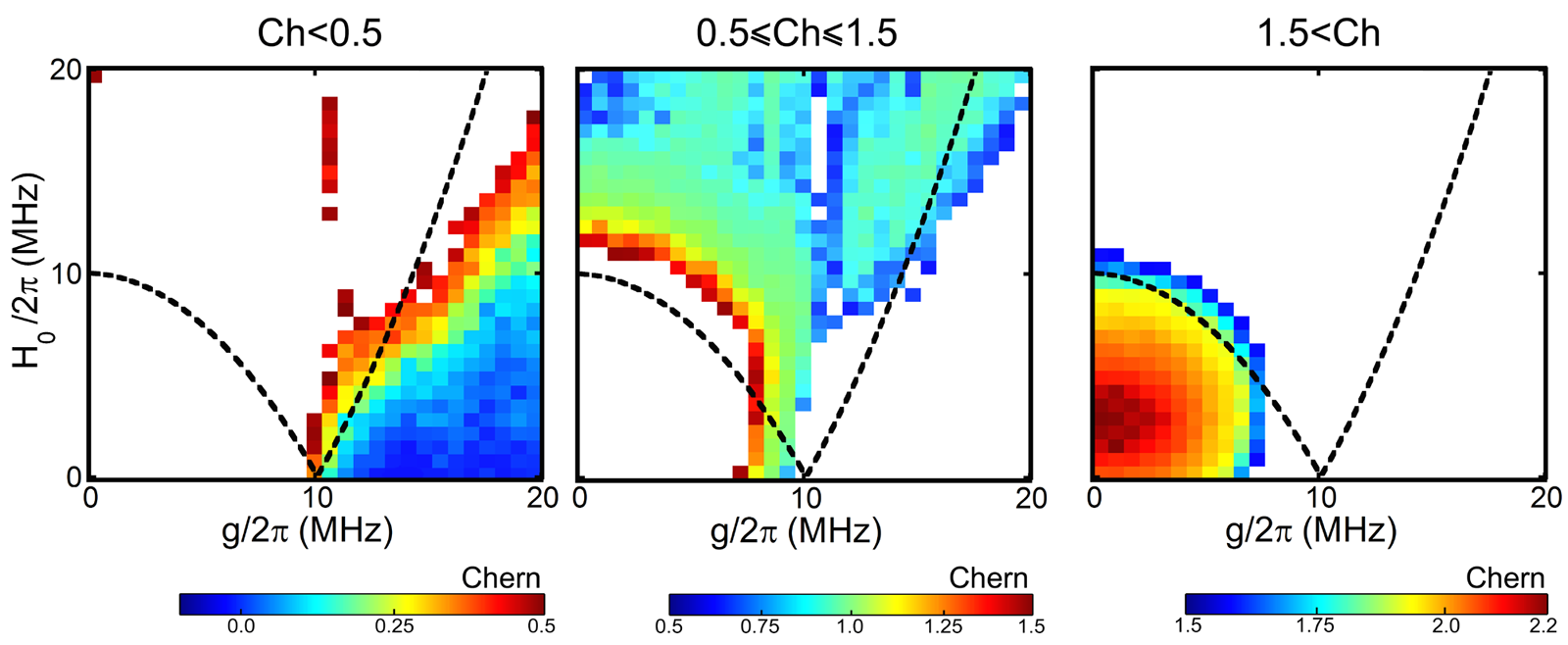}
\par\end{centering}
\caption{
\textbf{"Decomposition" of the topological phase diagram obtained with 2 qubits.} To demonstrate the fluctuation at each $\mathcal{C}\mathpzc{h}$ plateau and avoid obscuring it with the color map, here we replot the topological phase diagram shown in Fig. 4(c) of the main text (\(H_{r}/2\pi=10\,MHz\)). Each panel shows the data in a given interval of $\mathcal{C}\mathpzc{h}$ values. The same order of color tones is used in each panel, but the limits of the color scale for each panel is different. The black dashed lies are the fit using the analytical solution based on finding the loci of the monopoles(degeneracies of the Hamiltonian in this case). The deviations from the expected values have several sources: the crosstalk between the two qubits is likely the primary source, as the individual qubits were calibrated accurately. While the pulse length \(T_f\) was kept an order of magnitude smaller than the decoherence time in the system, decoherence and measurement errors also contribute to the error. Understanding these error mechanism is currently under way. The sharpness of the transition from one $\mathcal{C}\mathpzc{h}$ plateau to another is mainly related to the speed of the ramps. Slower ramping in parameter space (longer \(T_f\)) would result is sharper transitions. In order to successfully use slower ramps longer coherence times are required, which based on our current understanding of decoherence mechanisms gathered from this first generation of gmon devices, is achievable and will be implemented in the next generation of this experiment.}
\label{fig:concept}
\end{figure*}

\clearpage
\clearpage